\makeatletter
\@ifundefined{@parse@version@dash}{%
	\def\@parse@version#1{\@parse@version@0#1}
	\def\@parse@version@#1/#2/#3#4#5\@nil{%
		\@parse@version@dash#1-#2-#3#4\@nil}
	\def\@parse@version@dash#1-#2-#3#4#5\@nil{%
		\if\relax#2\relax\else#1\fi#2#3#4 }
}{}
\makeatother

\documentclass[aps,pre,twocolumn,groupedaddress,longbibliography]{revtex4-2}
\usepackage{amsfonts,amssymb,amsmath} 
\usepackage{color} 
\usepackage{graphicx} 
\usepackage{epstopdf,mathrsfs} 
\usepackage[colorlinks,linktocpage,linkcolor=blue]{hyperref}

\begin{document}

\title{Langevin picture of anomalous diffusion processes in expanding medium}

\author{Xudong Wang}
\author{Yao Chen}
\email{ychen@njau.edu.cn}
\affiliation{$^1$School of Mathematics and Statistics, Nanjing University of Science and Technology, Nanjing, 210094, P.R. China \\
$^2$College of Sciences, Nanjing Agricultural University, Nanjing, 210095, P.R. China}

\begin{abstract}
Expanding medium is very common in many different fields, such as biology and cosmology. It brings a nonnegligible influence on particle's diffusion, which is quite different from the effect of an external force field. The dynamic mechanism of particle's motion in expanding medium has only been investigated in the framework of continuous-time random walk. To focus on more diffusion processes and physical observables, we build the Langevin picture of anomalous diffusion in expanding medium, and conduct detailed analyses in the framework of Langevin equation. With the help of a subordinator, both subdiffusion process and superdiffusion process in expanding medium are discussed. We find that the expanding medium with different changing rate (exponential form and power-law form) leads to quite different diffusion phenomena. The particle's intrinsic diffusion behavior also plays an important role. Our detailed theoretical analyses and simulations present a panoramic view of investigating anomalous diffusion in expanding medium under the framework of Langevin equation.

\end{abstract}

\maketitle

\section{Introduction}\label{Sec1}
Beyond the classical Brownian motion, anomalous diffusion has become a very common phenomenon in the natural world recently. It is characterized by the nonlinear evolution of ensemble-averaged mean-squared displacement (EAMSD) with respect to time, i.e.,
\begin{equation}
  \langle x^2(t)\rangle \simeq 2D_\beta t^\beta
\end{equation}
with $\beta\neq1$ \cite{MetzlerKlafter:2000,BarkaiGariniMetzler:2012,HoflingFranosch:2013,Manzo:2015,MogreBrownKoslover:2020}.
One of the most typical model of describing particle's motion is continuous-time random walk (CTRW) \cite{MontrollWeiss:1965,KlafterSokolov:2011,ZaburdaevDenisovKlafter:2015}, which only consists of waiting time and jump length, but can describe many kinds of anomalous diffusion processes.
In CTRW, one typical example of subdiffusion with $\beta<1$ is characterized by power-law-distributed waiting times \cite{HeBurovMetzlerBarkai:2008,BurovJeonMetzlerBarkai:2011}. While superdiffusion examples with $\beta>1$ include L\'{e}vy flight with divergent second moment of jump length \cite{ShlesingerZaslavskyFrisch:1995,VahabiSchulzShokriMetzler:2013} and L\'{e}vy walk with heavy-tailed duration time of each running event \cite{TejedorMetzler:2010,MagdziarzMetzlerSzczotkaZebrowski:2012-1,ZaburdaevDenisovKlafter:2015,ChenWangDeng:2019}.

In recent years, it attracts people's attention that how particle diffuses in an expanding or contracting medium. Let $a(t)>0$ be the scale factor which describes the expanding rate of medium. For any fixed time $t$, $a(t)>1$ implies an expanding medium,
while $a(t)<1$ means a contracting medium.
For convenience, no matter $a(t)$ is greater or less than $1$, we call it the expanding medium for short in the following.
Some physical processes, especially particle's stochastic transport, are significantly influenced by the expanding or contracting effects of the medium. The expanding medium is common in the field of biology and cosmology. In biology, the examples include biological cells in interphase \cite{Alberts-etal:2015}, growing biological tissues \cite{Cowin:2004,Ambrosi-etal:2019} and lipid vesicles \cite{SzostakBartelLuisi:2001,XuHuChen:2016}. While In cosmology, the diffusion of cosmic rays in the expanding universe \cite{AloisioBerezinskyGazizov:2009} and the diffusion of fluids \cite{Haba:2014} are both worthy of investigation.

The dynamic mechanism of particle's motion in the expanding medium is quite different from that effected by an external force field, while the latter case has been discussed a lot for different kinds of force fields \cite{MagdziarzWeronKlafter:2008,EuleFriedrich:2009,CairoliBaule:2015,FedotovKorabel:2015,WangChenDeng:2020-2}. For the former case, in the framework of CTRW, the diffusing particles stick to the expanding medium and experience a drift even when they stay in the phase of waiting time. While at the moment of jumping event, the actual physical displacement is affected by the expanding rate $a(t)$ of medium. The Brownian motion \cite{YusteAbadEscudero:2016} and some common anomalous diffusion processes \cite{VotAbadYuste:2017,AngstmannHenryMcGann:2017,VotYuste:2018,Abad-etal:2020,VotAbadMetzlerYuste:2020} on evolving domain have been investigated to a certain extent. The corresponding Fokker-Planck equation can be derived to characterize the motion of diffusing particles on expanding medium, where the expanding rate $a(t)$ appears as the coefficient of drift term or diffusion term in macroscopic equations. The interplay between diffusive transport and the drift associated with the expanding medium gives rise to many striking effects, such as an enhanced memory of the initial condition \cite{YusteAbadEscudero:2016,VotAbadYuste:2017}, the slowing-down and even the premature halt of encounter-controlled reactions \cite{VotEscuderoAbadYuste:2018,EscuderoYusteAbadVot:2018}.

Despite these achievements on diffusion processes in expanding medium obtained in the framework of CTRW, there are still some missing of the important quantities, such as the particle's position correlation function and time-averaged mean-squared displacement (TAMSD).
Due to the rapid development of the technique of single particle tracking in studying transport processes in cellular membranes \cite{SaxtonJacobson:1997} and probe the microrheology of the cytoplasm \cite{Wirtz:2009,YaoTassieriPadgettCooper:2009}, TAMSD has become a useful observable by evaluating the particle's trajectory through video microscopy of fluorescently labeled molecules.
Compared with the probability density function (PDF) of particle's position and the EAMSD, the correlation function and TAMSD depend on the two-point joint PDF and contain more information of particle's trajectory. Another missing in CTRW framework is the discussion about L\'{e}vy walk in expanding medium. One possible difficulty is the coupling between the waiting time and jump length of L\'{e}vy walk.

In fact, as an alternative model of describing anomalous diffusion, Langevin equation has the advantage of including the effect of an external force field or noises generated from a fluctuating environment \cite{CoffeyKalmykovWaldron:2004}.
Fogedby \cite{Fogedby:1994} proposed that an overdamped Langevin equation in operation time $s$ coupled with a physical time process $t(s)$ (named as a subordinator) can model the same process as subdiffusive CTRW and L\'{e}vy flight in scaling limit. While an underdamped Langevin equation coupled with a subordinator can model the L\'{e}vy-walk-like diffusion processes in scaling limit \cite{EuleZaburdaevFriedrichGeisel:2012,WangChenDeng:2019}.

Based on these concerns, this paper aims at proposing a Langevin picture of describing anomalous diffusion processes in expanding medium.
Inspired by the fact that the Langevin equation can be regarded as the continuous time limit of a random walk model, based on the dynamic mechanism of particle's motion in expanding medium in CTRW framework, we build the Langevin equation of particle's physical coordinate in expanding medium, and then conduct the detailed analyses.

The remainder of this paper is organized as follows. In Sec. \ref{Sec2}, we show the detailed mathematic description of the dynamic mechanism of particle's motion in expanding medium under the framework of both CTRW model and Langevin equation, together with some elementary knowledge of a subordinator. Then we investigate the specific subdiffusion and superdiffusion processes in expanding medium under the framework of Langevin equation in Secs. \ref{Sec3} and \ref{Sec4}, respectively.
A summary of the main results is provided in Sec. \ref{Sec5}. In the appendices some mathematical details are collected.

\section{Expanding medium models and subordinator}\label{Sec2}

\begin{figure}
  \centering
  \includegraphics[scale=0.6]{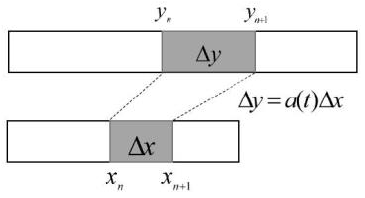}\\
  \caption{Schematic representation of the relationship between physical coordinate $y(t)$ and comoving coordinate $x(t)$ at time $t$. The upper rectangle denotes the expanding medium at time $t$, while the lower one the original medium at the initial time.}\label{fig0}
\end{figure}

The particle's motion on expanding medium has been formulated in CTRW model \cite{VotAbadYuste:2017,AngstmannHenryMcGann:2017,VotYuste:2018,Abad-etal:2020,VotAbadMetzlerYuste:2020}. The CTRW model consists of a series of waiting times and jump lengths, where the particle stays in some certain position for a random time $\Delta t$ at one waiting state and performs the instantaneous jump with random length $\Delta y$. For a particle on expanding medium, its physical position changes even at the waiting states as the medium expands. For a clear description of the change of particle's physical coordinate $y(t)$, a new comoving coordinate $x(t)$, which is associated with a reference frame where the expanding medium appears to be static, is introduced in Ref. \cite{VotAbadYuste:2017}. The two coordinates are related through an equality for any physical time $t$:
\begin{equation}\label{Eq-at1}
  y(t)=a(t)x(t),
\end{equation}
where $a(t)$ is the so-called scale factor satisfying $a(0)=1$.
The advantage of introducing comoving coordinate is that the descriptions of both waiting states and jumping states are effective with respect to the comoving coordinate. More precisely, at waiting states, the particle keeps still with respect to the expanding medium itself, in other words, its comoving coordinate $x(t)$ does not change. While at jumping states, letting the jump length PDF $w_y(\Delta y)$ describe the intrinsic random motion of the particle, the corresponding jump length with respect to the comoving coordinate should be
\begin{equation}\label{Eq-at2}
  \Delta x=\Delta y/a(t),
\end{equation}
where $t$ is the instant when the jump happens. See the graph in Fig. \ref{fig0}, where the $n$-th step in physical and comoving coordinates are $\Delta y$ and $\Delta x$, respectively.
Therefore, the jump length PDF $w_x(\Delta x)$ with respect to the comoving coordinate is
\begin{equation}
  w_x(\Delta x)=a(t)w_y(a(t)\Delta x),
\end{equation}
becoming time-dependent. Knowing the waiting time and jump length PDFs, the standard procedures of CTRW model can be applied to obtain the information of comoving coordinate $x(t)$. Thus, the physical coordinate $y(t)$ can be obtained by use of the relation in Eq. \eqref{Eq-at1}.

Our aim is to extend the procedures in CTRW framework to Langevin equation, which describes the particle's motion by a stochastic differential equation of physical coordinate $y(t)$.
In Langevin equation, the particle's trajectory can be approximated by the cumulative sum of increments $\Delta y$ in all different time intervals $[t,t+\Delta t]$. In this approximation, the particle's motion in time interval $[t,t+\Delta t]$ can be understood as the combination of a waiting time $\Delta t$ and a jump length $\Delta y$. Therefore, similar to the method used in CTRW framework, the comoving coordinate $x(t)$ can be introduced in Langevin equation, and the relations in Eqs. \eqref{Eq-at1} and \eqref{Eq-at2} are also valid. Then dividing $\Delta t$ on both sides of Eq. \eqref{Eq-at2} and letting $\Delta t\rightarrow0$, we find that the Langevin equations of comoving coordinate $x(t)$ and intrinsic displacement $y_I(t)$ only differs by the scale factor $a(t)$, i.e.,
\begin{equation}\label{Eq-at3}
  \dot{x}(t)=\dot{y}_I(t)/a(t).
\end{equation}
Note that the intrinsic displacement $y_I(t)=\sum_{i}\Delta y_i$ denotes the particle's position at time $t$ without an expanding medium, which is different from the physical coordinate $y(t)$ in Eq. \eqref{Eq-at1} for the case with an expanding medium.

Although the difference is only the scale factor $a(t)$ in Eq. \eqref{Eq-at3}, the analyses of the diffusion behavior of comoving coordinate $x(t)$ are not trivial, especially when the Langevin equation is coupled with a subordinator $t(s)$. Subordinator is a non-decreasing L\'{e}vy process \cite{Applebaum:2009} and can be regarded as a stochastic model of time evolution. In coupled Langevin equation, the particle's physical coordinate is denoted as a compound process $y(t):=y(s(t))$, where $s(t)$ in the corresponding inverse subordinator, defined by
\begin{equation}
  s(t)=\inf_{s>0}\{s:t(s)>t\}.
\end{equation}
There are two time variables in the coupled Langevin equation, physical time $t$ and operational time $s$. Nevertheless, Eq. \eqref{Eq-at3} with scale factor $a(t)$ is only valid with respect to physical time $t$.

The subordinator has been commonly used in Langevin system to describe different kinds of subdiffusion \cite{Fogedby:1994,MetzlerKlafter:2000-2,MetzlerKlafter:2000-3,ChenWangDeng:2018-2} and superdiffusion \cite{FriedrichJenkoBauleEule:2006,FriedrichJenkoBauleEule:2006-2,EuleZaburdaevFriedrichGeisel:2012,WangChenDeng:2019,ChenWangDeng:2019-3,ChenWang:2021} processes when coupled with an overdamped and underdamped Langevin equation, respectively.
In order to characterize the power-law-distributed waiting time in CTRW model, the subordinator $t(s)$ in this paper is taken to be $\alpha$-dependent ($0<\alpha<2$) with its characteristic function being
\begin{equation}\label{char}
  g(\lambda,s):=\langle e^{-\lambda t(s)}\rangle=e^{-s\Phi(\lambda)},
\end{equation}
where the Laplace exponent \cite{BauleFriedrich:2005,WangChenDeng:2019,ChenWangDeng:2019-3} is
\begin{equation}\label{Phi}
\begin{split}
\Phi(\lambda)=\left\{
\begin{array}{ll}
 \lambda^\alpha,  & ~0<\alpha<1,  \\[5pt]
  \frac{\tau_0}{\alpha-1}\lambda-\tau^\alpha_0|\Gamma(1-\alpha)| \lambda^\alpha, &~ 1<\alpha<2.
\end{array}\right.
\end{split}
\end{equation}
The two-point PDF of the subordinator $t(s)$ can be expressed as
\begin{equation}\label{g}
  g(t_1,s_1;t_2,s_2)= \langle \delta(t_1-t(s_1)) \delta(t_2-t(s_2)) \rangle.
\end{equation}
By virtue of the stationary and independent increments of subordinator $t(s)$, the two-point PDF $h(s_2,t_2;s_1,t_1)$ of the inverse subordinator $s(t)$ has the expression in Laplace space $(t_1\rightarrow\lambda_1,t_2\rightarrow\lambda_2)$ \cite{BauleFriedrich:2005}
\begin{equation}\label{h2}
\begin{split}
&h(s_1,\lambda_1;s_2,\lambda_2)  \\
    &=\frac{\partial}{\partial s_1} \frac{\partial}{\partial s_2} \frac{1}{\lambda_1\lambda_2}\,g(\lambda_1,s_1;\lambda_2,s_2) \\
    &=\delta(s_2-s_1)\frac{\Phi(\lambda_1)+\Phi(\lambda_2)-\Phi(\lambda_1+\lambda_2)}{\lambda_1\lambda_2}\,e^{-s_1\Phi(\lambda_1+\lambda_2)} \\
    &~~~+\Theta(s_2-s_1)\frac{\Phi(\lambda_2)(\Phi(\lambda_1+\lambda_2)-\Phi(\lambda_2))}{\lambda_1\lambda_2} \\
    &~~~\times e^{-s_1\Phi(\lambda_1+\lambda_2)}e^{-(s_2-s_1)\Phi(\lambda_2)}   \\
    &~~~+\Theta(s_1-s_2)\frac{\Phi(\lambda_1)(\Phi(\lambda_1+\lambda_2)-\Phi(\lambda_1))}{\lambda_1\lambda_2} \\
    &~~~\times e^{-s_2\Phi(\lambda_1+\lambda_2)}e^{-(s_1-s_2)\Phi(\lambda_1)}.
\end{split}
\end{equation}
The normalization of $h(s_2,\lambda_2;s_1,\lambda_1)$ can be verified through the equality $\int_0^\infty \int_0^\infty  h(s_2,\lambda_2;s_1,\lambda_1)ds_1 ds_2=(\lambda_1\lambda_2)^{-1}$. The two-point PDF $h(s_2,t_2;s_1,t_1)$ will be used to evaluate the correlation function in physical time $t$.
The dynamic behaviors of particles moving in expanding medium will be investigated for both subdiffusion and superdiffusion cases.
The subdiffusion case only uses the range $0<\alpha<1$ in the subordinator, while the superdiffusion case considers both $0<\alpha<1$ and $1<\alpha<2$ due to its richer diffusion behaviors \cite{BauleFriedrich:2005,WangChenDeng:2019,ChenWangDeng:2019-3}.

\section{Subdiffusion in expanding medium}\label{Sec3}

Let us first consider a subdiffusion process described by an overdamped Langevin equation coupled with a subordinator \cite{Fogedby:1994,BauleFriedrich:2005,BauleFriedrich:2007,Magdziarz:2009,ChechkinSokolov:2021}
\begin{equation}\label{model1}
\dot{y}_I(s)=\sqrt{2D}\xi(s), \quad \dot{t}(s)=\eta(s),
\end{equation}
where $D$ is the constant diffusivity, $\xi(s)$ is the Gaussian white noise with zero mean value and correlation function $\langle \xi(s_1)\xi(s_2) \rangle=\delta(s_1-s_2)$, $t(s)$ is the $\alpha$-stable subordinator ($0<\alpha<1$) with the characteristic function in Eq. \eqref{char}. The notation $y_I(s)$ denotes the particle's intrinsic displacement over operational time $s$ without an expanding medium.  The L\'{e}vy noise $\eta(s)$, regarded as the formal derivative of the $\alpha$-stable subordinator $t(s)$, is independent of the Gaussian white noise $\xi(s)$.

Equation \eqref{model1} describes the intrinsic random motion of particles without considering an expanding medium. The key of considering the effect of the expanding medium is to build the Langevin equation of comoving coordinate $x(t)$, which satisfies the relation in Eq. \eqref{Eq-at3}. For this purpose, the two sub-equations in Eq. \eqref{model1} should be merged into one equation of physical coordinate $y_I(t)$, which is
\begin{equation}\label{modelyt}
  \dot{y}_I(t)=\sqrt{2D}\bar\xi(t),
\end{equation}
where the new noise $\bar\xi(t)$ is defined as \cite{CairoliBaule:2015-2,ChenWangDeng:2019-2}
\begin{equation}\label{Def-newnoise}
  \bar\xi(t):=\frac{dB(s(t))}{dt}=\xi(s(t))\frac{ds(t)}{dt},
\end{equation}
and $B(\cdot)$ is the standard Brownian motion.
Equation \eqref{modelyt} is obtained by replacing $s$ by the inverse subordinator $s(t)$ in the first equation of Eq. \eqref{model1} and using the definition of compound process $y_I(t):=y_I(s(t))$.
Therefore, for the subdiffusion process in expanding medium with scale factor $a(t)$, the Langevin equation with respect to the comoving coordinate $x(t)$ is
\begin{equation}\label{modelxt}
\begin{split}
\dot{x}(t)&=\frac{\sqrt{2D}}{a(t)}\bar\xi(t).
\end{split}
\end{equation}
The corresponding Fokker-Planck equation governing the PDF $W(x,t)$ of finding the particle's comoving coordinate $x$ at time $t$, can be derived from Langevin equation \eqref{modelxt} by using the common method in Refs. \cite{SokolovKlafter:2006,SokolovKlafter:2006,MagdziarzWeronKlafter:2008,Magdziarz:2009,EuleFriedrich:2009,CairoliBaule:2017,ChenWangDeng:2019-2}
\begin{equation}\label{FP}
\begin{split}
\frac{\partial W(x,t)}{\partial t}=\frac{2D}{a^2 (t)}\frac{\partial^2}{\partial x^2}[D_t^{1-\alpha}W(x,t)].
\end{split}
\end{equation}
The most direct way of obtaining Eq. \eqref{FP} is taking the Laplace symbol $p=0$ in Feynman-Kac equation (71) of Ref. \cite{CairoliBaule:2017}. The consistence between Eq. \eqref{FP} and Eq. (61) of Ref. \cite{VotAbadYuste:2017} derived under the CTRW framework implies that the Langevin picture of subdiffusion process in expanding medium is effective.

The moments of the comoving coordinate $x(t)$ and the shape of PDF $W(x,t)$ can be obtained by use of Eq. \eqref{FP} as Ref. \cite{VotAbadYuste:2017} shows. Moreover, based on the Langevin equation \eqref{modelxt}, more quantities, such as the position correlation function and TAMSD, can be investigated.
In detail, the comoving coordinate $x(t)$ can be solved from the Langevin equation \eqref{modelxt}, i.e.,
\begin{equation}\label{model1-xt}
\begin{split}
x(t)=\sqrt{2D}\int_0^t\frac{\bar\xi(t')}{a(t')}dt'.
\end{split}
\end{equation}
Then considering the correlation function of the noise $\bar\xi(t)$ (see Appendix \ref{App2} for the derivation or refer to Refs. \cite{CairoliBaule:2015-2,ChenWangDeng:2019-2})
\begin{equation}\label{corr-newxi}
  \langle \bar\xi(t_1)\bar\xi(t_2)\rangle=t_1^{\alpha-1}\delta(t_1-t_2)/\Gamma(\alpha),
\end{equation}
the EAMSD of the subdiffusion process in comoving coordinate $x(t)$ is
\begin{equation}\label{x2}
\begin{split}
\langle x^2(t)\rangle&=2D\int_0^t\int_0^t\frac{\langle \bar\xi(t_1')\bar\xi(t_2')\rangle}{a(t_1')a(t_2')}dt_1'dt_2' \\
&=\frac{2D}{\Gamma(\alpha)}\int_0^t  \frac{t'^{\alpha-1}}{a^2(t')}dt'.
\end{split}
\end{equation}
Due to the $\delta$-correlation of noise $\bar\xi(t)$ in Eq. \eqref{corr-newxi}, the correlation function of comoving correlation $x(t)$ satisfies
\begin{equation}
\begin{split}
\langle x(t_1)x(t_2)\rangle=\langle x^2(t_1)\rangle
\end{split}
\end{equation}
for $t_1<t_2$. Based on the quantities over comoving coordinate $x(t)$, those over physical coordinate $y(t)$ can be obtained by use of Eq. \eqref{Eq-at1}, which are the EAMSD
\begin{equation}\label{EAMSD1}
\begin{split}
\langle y^2(t)\rangle=a^2(t)\langle x^2(t)\rangle=\frac{2Da^2(t)}{\Gamma(\alpha)}\int_0^t  \frac{t'^{\alpha-1}}{a^2(t')}dt',
\end{split}
\end{equation}
and the correlation function for $t_1<t_2$
\begin{equation}
\begin{split}
\langle y(t_1)y(t_2)\rangle&=a(t_1)a(t_2)\langle x^2(t_1)\rangle\\
&=\frac{a(t_2)}{a(t_1)}\langle y^2(t_1)\rangle,
\end{split}
\end{equation}
respectively. Thus, based on the definition of TAMSD \cite{MetzlerJeonCherstvyBarkai:2014,BurovJeonMetzlerBarkai:2011}
\begin{equation}\label{Def-TAMSD}
  \overline{\delta^2(\Delta)}=\frac{1}{T-\Delta}\int_{0}^{T-\Delta} [x(t+\Delta)-x(t)]^2 dt,
\end{equation}
we obtain the ensemble-averaged TAMSD
\begin{equation}\label{TAMSD}
\begin{split}
  \langle\overline{\delta^2(\Delta)}\rangle
  &=\frac{1}{T-\Delta}\int_0^{T-\Delta} \langle y^2(t+\Delta)\rangle\\
  &~~~+\left(1-2\frac{a(t+\Delta)}{a(t)}\right)\langle y^2(t)\rangle dt.
\end{split}
\end{equation}

The explicit dependence of the quantities on the scale factor $a(t)$ implies that the particle's diffusion behavior is indeed affected by the expanding medium. The results on the physical coordinate in Eqs. \eqref{EAMSD1}-\eqref{TAMSD} are valid for any form of $a(t)$. By taking $a(t)=1$, these results recover to those of the subdiffusion process in a static medium. For example in Eq. \eqref{EAMSD1}, $a(t)=1$ yields the subdiffusion behavior with $\langle y^2(t)\rangle=2Dt^\alpha/\Gamma(\alpha+1)$. In the expanding medium, however, the scale factor $a(t)$ appears both inside and outside the integral of Eq. \eqref{EAMSD1}. The one in the denominator of integral is resulted from Eq. \eqref{Eq-at2}, which transforms the particle's intrinsic motion into comoving coordinate. While another one appears outside the integral, which is yielded by Eq. \eqref{Eq-at1} and turns the comoving coordinate to the physical coordinate. Since $a(t)=1$ when $t=0$, the short time limit of the quantities will be the same as the case without an expanding medium. We mainly consider the long time limit for cases with different scale factor $a(t)$ in the following.

\subsection{Scale factor $a(t)=e^{\gamma t}$}\label{harmonic}

When the medium changes exponentially with scale factor $a(t)=e^{\gamma t}$, one has
\begin{equation}\label{shou}
\begin{split}
\langle x^2(t)\rangle=\frac{2Dt^\alpha}{\Gamma(\alpha+1)} {}_1F_1(\alpha,\alpha+1;-2\gamma t),
\end{split}
\end{equation}
where the confluent hypergeometric function has the definition and asymptotic expression for large $z$ \cite{AbramowitzStegun:1972}
\begin{equation}
  \begin{split}
    _1F_1(a, b; z)&=\frac{\Gamma(b)}{\Gamma(a)\Gamma(b-a)}\int_0^1e^{zu}u^{a-1}(1-u)^{b-a-1}du \\
    &\simeq \Gamma(b)\left( e^z z^{a-b}/\Gamma(a)+(-z)^{-a}/\Gamma(b-a)\right).
  \end{split}
\end{equation}
Therefore, the asymptotic EAMSD on comoving coordinate is
\begin{equation}
  \begin{split}
    \langle x^2(t)\rangle\simeq\left\{
    \begin{array}{ll}
          2^{1-\alpha}\gamma^{-\alpha}D, &~ \gamma>0, \\[5pt]
      \frac{D}{|\gamma|\Gamma(\alpha)}t^{\alpha-1}e^{2|\gamma| t}, & ~\gamma<0.
    \end{array}\right.
  \end{split}
\end{equation}
Considering the relation $y(t)=a(t)x(t)$ between two kinds of coordinates, the EAMSD on physical coordinate can be obtained
\begin{equation}\label{EAMSD1-y}
  \begin{split}
    \langle y^2(t)\rangle\simeq\left\{
    \begin{array}{ll}
          2^{1-\alpha}\gamma^{-\alpha}De^{2\gamma t}, &~ \gamma>0, \\[5pt]
      \frac{D}{|\gamma|\Gamma(\alpha)}t^{\alpha-1}, & ~\gamma<0,
    \end{array}\right.
  \end{split}
\end{equation}
which is consistent to the results in Ref. \cite{VotAbadYuste:2017}.
The exponentially expanding medium with $\gamma>0$ yields a superdiffusion behavior of exponential form, while the exponentially contracting medium with $\gamma<0$ leads to a power-law decaying of EAMSD due to $\alpha<1$. For the former case with $\gamma>0$, the particle's intrinsic motion is negligible compared with the exponential expanding of medium. So the EAMSD of comoving coordinate $\langle x^2(t)\rangle$ converges to a constant at long time limit, while the corresponding $\langle y^2(t)\rangle$ increases at an exponential rate as the medium itself.
The latter case with $\gamma<0$ presents the same diffusion behavior as the subdiffusion process in harmonic potential \cite{ChenWangDeng:2019-2}, where the external force acts on the subordinated process $y(t)$ and drags the particle towards zero for all physical times.

In fact, the effect of the exponentially contracting medium with $\gamma<0$ on the subdiffusion process is equivalent to that of a harmonic potential $U(y)=|\gamma|y^2/2$, which can be justified by comparing the Langevin equation of physical coordinate $y(t)$ in the two cases. For a general process driven by random noise $\zeta(t)$, the Langevin equation of $y(t)$ containing a harmonic potential (or an external force $U'(y)=-|\gamma|y$) is
\begin{equation}\label{y_harmonic}
\begin{split}
\frac{dy(t)}{dt}&=-|\gamma| y(t)+\sqrt{2D}\zeta(t),
\end{split}
\end{equation}
the solution of which is
\begin{equation}\label{solution}
  y(t)=\sqrt{2D}\int_0^{t}e^{-|\gamma| (t-\tau)}\zeta(\tau)d\tau,
\end{equation}
with the initial condition $y_0=0$. On the other hand, by substituting $a(t)=e^{\gamma t}$ into Eq. \eqref{model1-xt}, replacing $\bar\xi(t)$ by the noise $\zeta(t)$, and considering the relation $y(t)=a(t)x(t)$, one can arrive at the same expression as Eq. \eqref{solution} for the particle moving in an exponentially contracting medium. The equivalent Langevin equation implies that the idea of considering the particle's motion in an exponentially contracting medium is an alternative way of investigating the diffusion behavior of the particle affected by a harmonic potential. The equivalence is valid for arbitrary random noise $\zeta(t)$, which means that the idea can be applied to a large amount of anomalous diffusion processes.

As for the ensemble-averaged TAMSD of particles, the large-$t$ behavior of the integrand in Eq. \eqref{TAMSD} plays the dominating role due to the precondition $\Delta\ll T$ \cite{MeyerBarkaiKantz:2017,WangChen:2022}. Therefore, we substitute the asymptotic EAMSD in Eq. \eqref{EAMSD1-y} into Eq. \eqref{TAMSD}, and obtain
\begin{equation}\label{EATAMSD1-y}
  \begin{split}
    \langle\overline{\delta^2(\Delta)}\rangle\simeq\left\{
    \begin{array}{ll}
         \frac{4D}{(2\gamma)^{1+\alpha}}T^{-1}e^{2\gamma T}, &~ \gamma>0, \\[5pt]
      \frac{2D}{|\gamma|\Gamma(\alpha+1)} T^{\alpha-1}, & ~\gamma<0,
    \end{array}\right.
  \end{split}
\end{equation}
for large lag time $\Delta$, where the latter result with $\gamma<0$ is consistent to that of the subdiffusion process in a harmonic potential \cite{ChenWangDeng:2019-2}. This consistence also verifies the equivalence between the effect of an exponentially contracting medium and a harmonic potential.
The ensemble-averaged TAMSD tends to be independent of lag time $\Delta$ for both $\gamma>0$ and $\gamma<0$, which shows obvious difference from the EAMSD is Eq. \eqref{EAMSD1-y}. Therefore, the Langevin system containing an expanding medium with exponential scale factor $a(t)$ is nonergodic. Even when $\alpha=1$, the subdiffusion process returns to the classical Brownian motion, the ensemble-averaged TAMSD is not equal to the corresponding EAMSD, either. Note that for $\gamma<0$, similar to a harmonic potential, the ensemble-averaged TAMSD tends to be twice the EAMSD for Browian motion with $\alpha=1$, i.e.,
\begin{equation}
  \langle\overline{\delta^2(\Delta)}\rangle\simeq \frac{2D}{|\gamma|}\simeq 2\langle y^2(\Delta)\rangle,
\end{equation}
which can also be found in other anomalous diffusion processes \cite{JeonMetzler:2012,WangChenDeng:2020-2}.

\begin{figure}
\begin{minipage}{0.35\linewidth}
  \centerline{\includegraphics[scale=0.29]{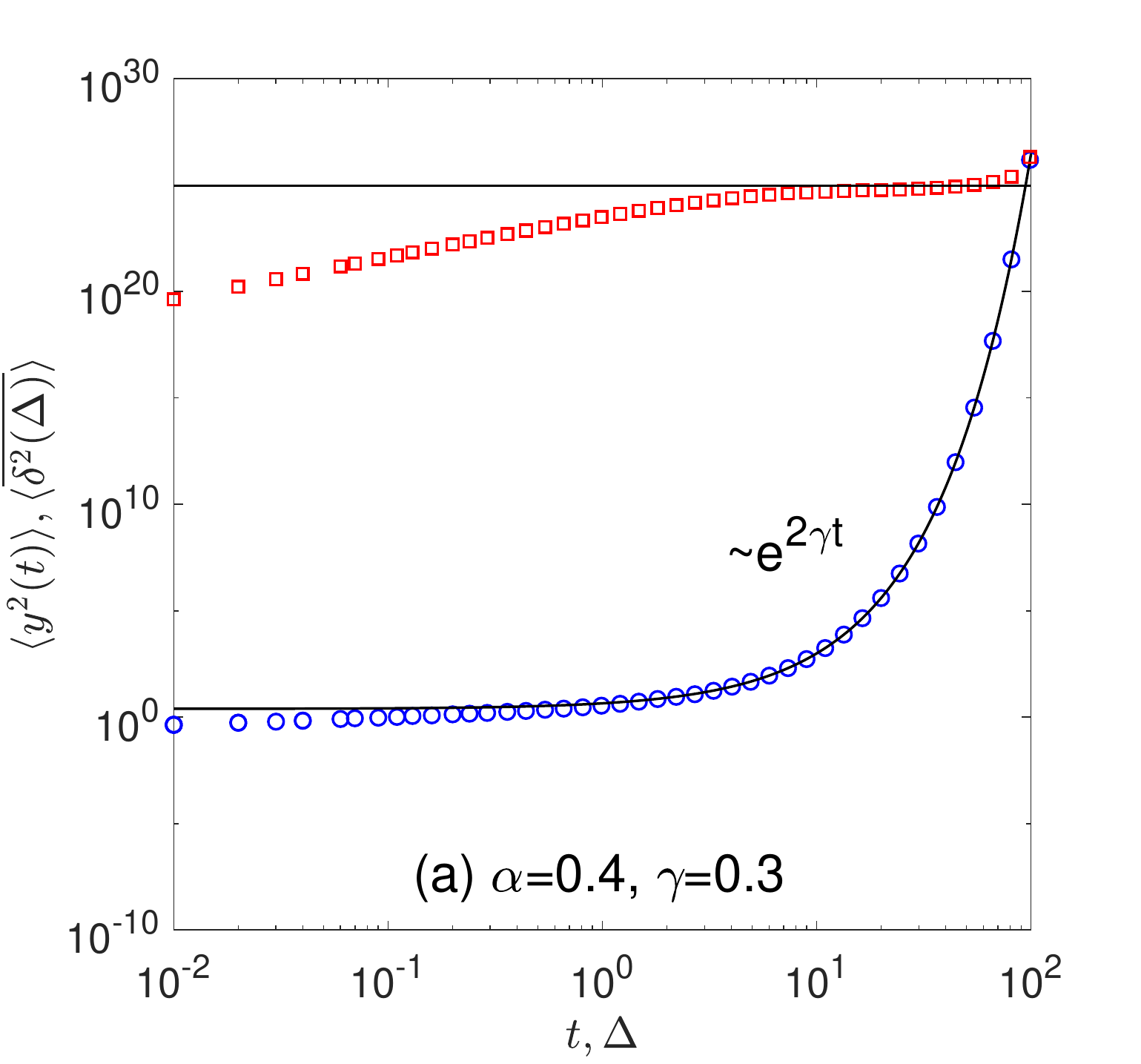}}
  \centerline{}
\end{minipage}
\hspace{1.43cm}
\begin{minipage}{0.35\linewidth}
  \centerline{\includegraphics[scale=0.29]{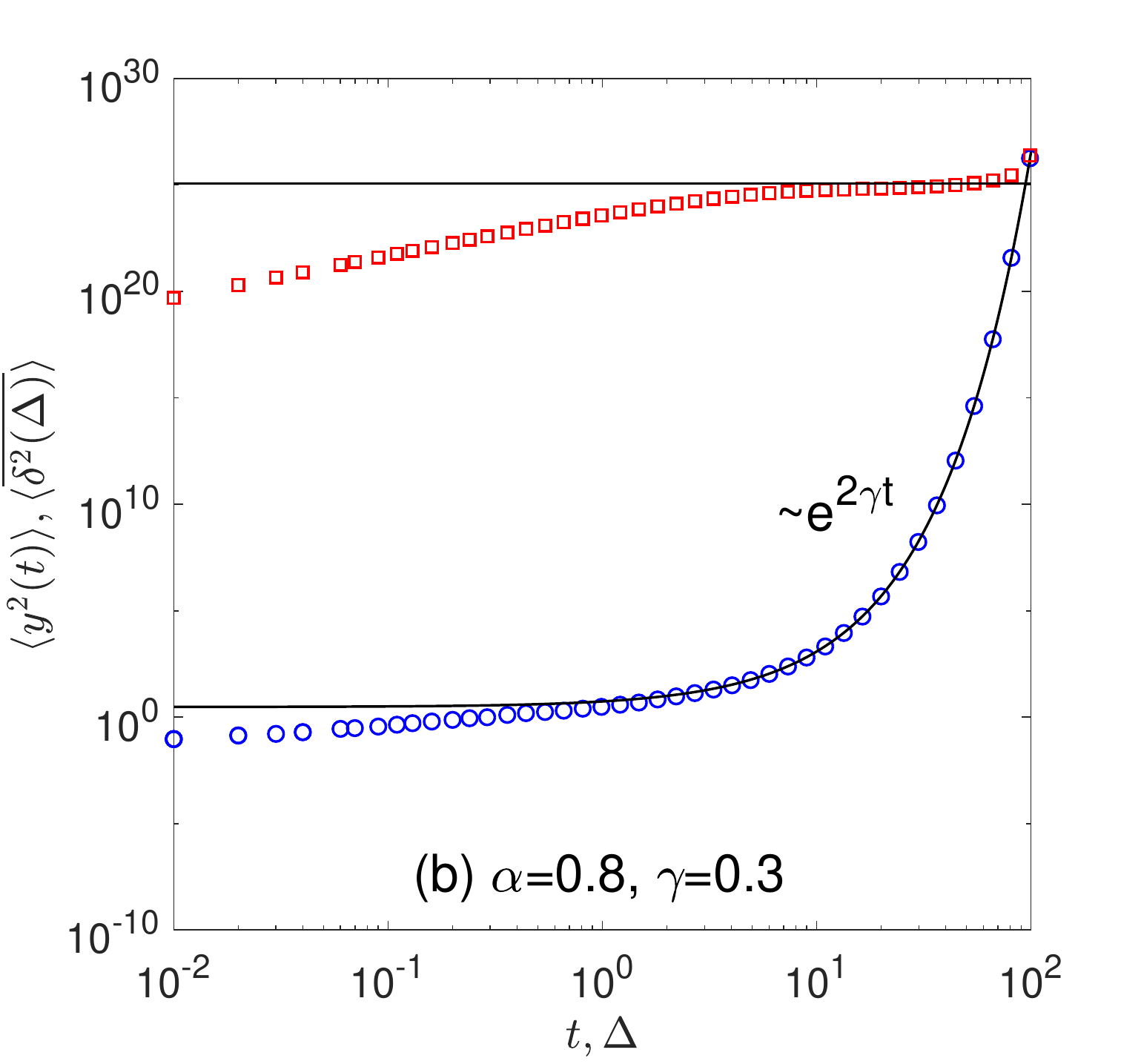}}
  \centerline{}
\end{minipage}
\begin{minipage}{0.35\linewidth}
  \centerline{\includegraphics[scale=0.29]{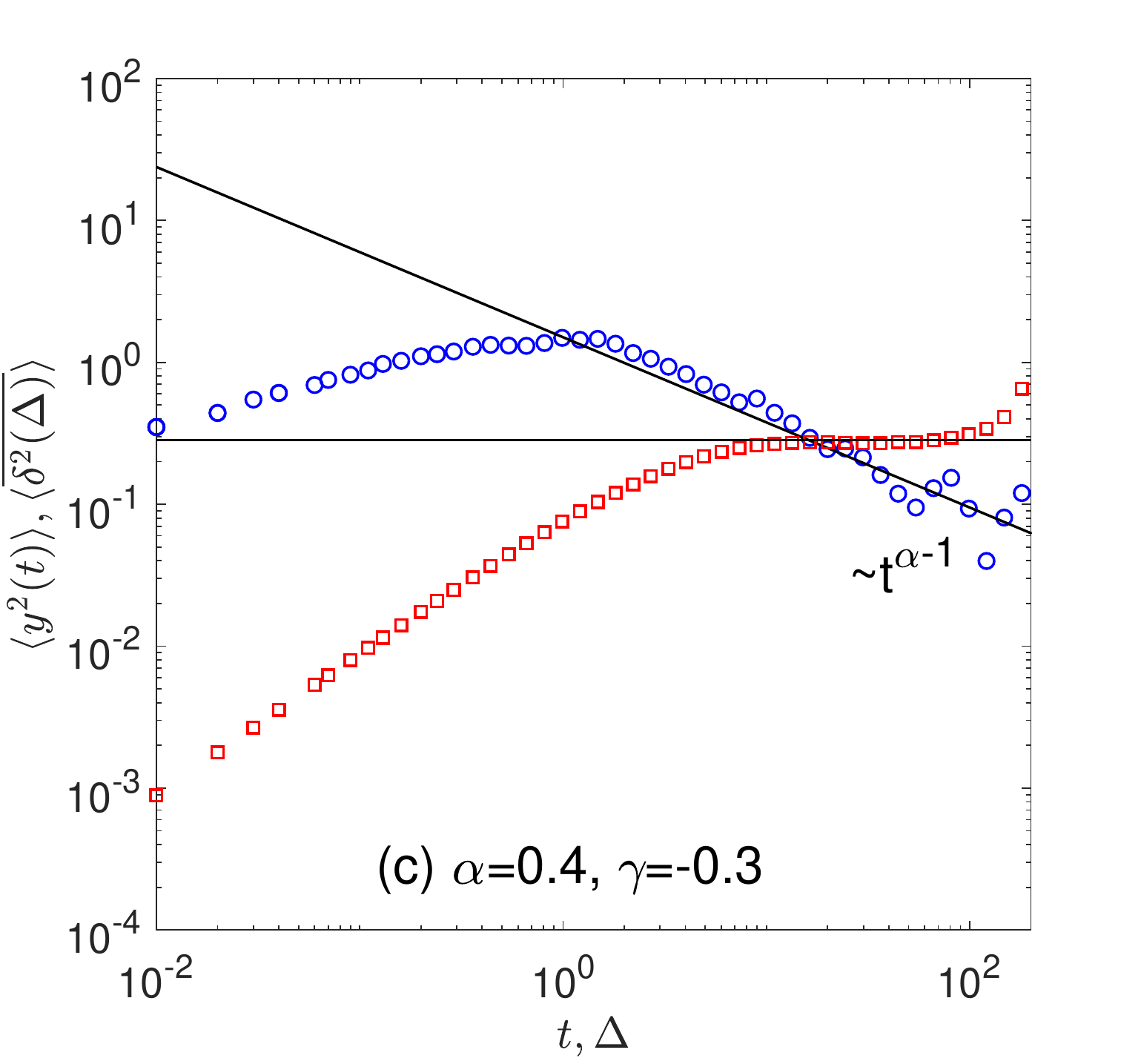}}
  \centerline{}
\end{minipage}
\hspace{1.43cm}
\begin{minipage}{0.35\linewidth}
  \centerline{\includegraphics[scale=0.29]{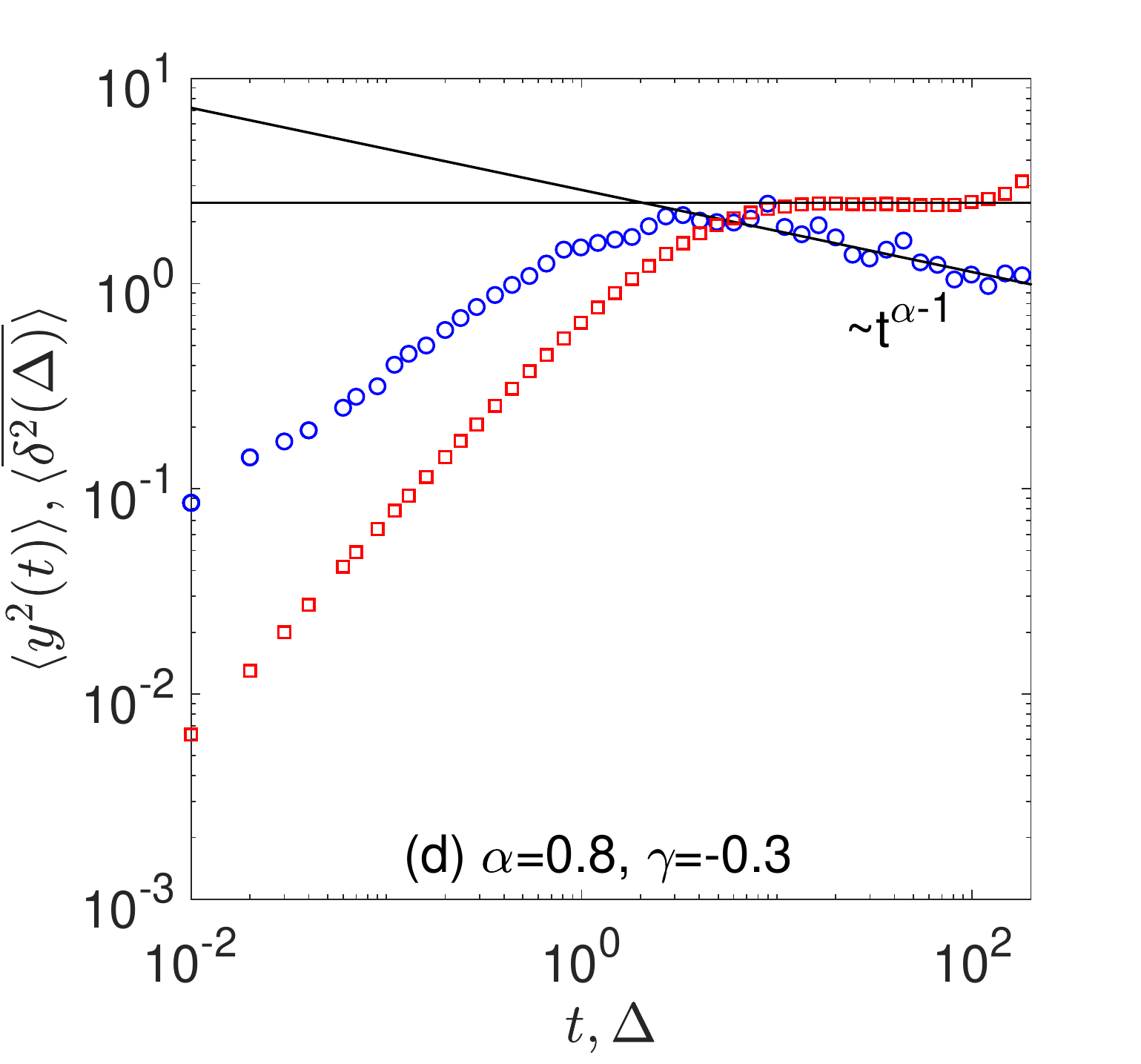}}
  \centerline{}
\end{minipage}
\caption{(Color online) EAMSD $\langle x^2(t)\rangle$ and ensemble-averaged TAMSD $\langle\overline{\delta^2(\Delta)}\rangle$ when the medium changes exponentially with scale factor $a(t)=e^{\gamma t}$. The theoretical results for $\langle x^2(t)\rangle$ in Eq. \eqref{EAMSD1-y} and $\langle\overline{\delta^2(\Delta)}\rangle$ in Eq. \eqref{EATAMSD1-y} are shown by black solid lines. The blue circles and red squares denote the simulated EAMSD and ensemble-averaged TAMSD, respectively. They fit to the theoretical lines well in four panels with different $\alpha$ and $\gamma$. The last several simulation points for ensemble-averaged TAMSD deviate slightly from theoretical results due to the failure of condition $\Delta\ll T$.
Other parameters: $D=1$, $T=100$ in (a-b) and $T=200$ in (c-d), and the number of trajectories used for ensemble is $N=10^3$.}\label{fig1}
\end{figure}

The simulations of EAMSD and ensemble-averaged TAMSD are presented in Fig. \ref{fig1}, where we choose different $\alpha$ and $\gamma$. The positive $\gamma$ in the upper panel of Fig. \ref{fig1} implies an exponentially expanding medium, while the negative one in the bottom panel implies an exponentially contracting medium. The EAMSD is sensitively affected by the medium, presenting exponential growth and power-law decaying for expanding and contracting medium, respectively.
By contrast, the ensemble-averaged TAMSD tends to a constant independent of the lag time $\Delta$ for both of the two kinds of media as Eq. \eqref{EATAMSD1-y} shows.

\subsection{Scale factor $a(t)=(\frac{t+t_0}{t_0})^\gamma$}
When the medium changes in a power-law rate with scale factor $a(t)=(\frac{t+t_0}{t_0})^\gamma$, one has
\begin{equation}\label{MSD-1Bx1}
\begin{split}
\langle x^2(t)\rangle=\frac{2D}{\Gamma(1+\alpha)}t^\alpha {_2F_1\left(2\gamma,\alpha;1+\alpha;-\frac{t}{t_0}\right)},
\end{split}
\end{equation}
where the Gaussian hypergeometric function has the definition for $c>b>0$ and asymptotic expression for large $z$ \cite{AbramowitzStegun:1972}
\begin{equation}\label{F21}
  \begin{split}
    _2F_1&(a, b;c; z) \\
    &=\frac{\Gamma(c)}{\Gamma(b)\Gamma(c-b)}\int_0^1(1-zu)^{-a}u^{b-1}(1-u)^{c-b-1}du \\
    &\simeq  \frac{\Gamma(c)\Gamma(b-a)}{\Gamma(b)\Gamma(c-a)}(-z)^{-a}.
  \end{split}
\end{equation}
Note that the asymptotic expression above is only valid for $a<b$. Otherwise, exchanging $a$ and $b$ will yield the correct result.
Therefore, for the EAMSD on comoving coordinate, it holds that
\begin{equation}\label{MSD-1Bx2}
  \begin{split}
    \langle x^2(t)\rangle\simeq\left\{
    \begin{array}{ll}
          \frac{2Dt_0^{\alpha}\Gamma(2\gamma-\alpha)}{\Gamma(2\gamma)}, &~ \gamma>\frac{\alpha}{2}, \\[5pt]
      \frac{2Dt_0^{2\gamma}}{(\alpha-2\gamma)\Gamma(\alpha)}t^{\alpha-2\gamma}, & ~\gamma<\frac{\alpha}{2}.
    \end{array}\right.
  \end{split}
\end{equation}
Considering the relation $y(t)=a(t)x(t)$ between two kinds of coordinates, the EAMSD on physical coordinate can be obtained
\begin{equation}\label{MSD-1By}
  \begin{split}
    \langle y^2(t)\rangle\simeq\left\{
    \begin{array}{ll}
          \frac{2Dt_0^{\alpha-2\gamma}\Gamma(2\gamma-\alpha)}{\Gamma(2\gamma)}t^{2\gamma}, &~ \gamma>\frac{\alpha}{2}, \\[5pt]
      \frac{2D}{(\alpha-2\gamma)\Gamma(\alpha)}t^{\alpha}, & ~\gamma<\frac{\alpha}{2}.
    \end{array}\right.
  \end{split}
\end{equation}
Note that the latter case $\gamma<\alpha/2$ is valid for both $\gamma>0$ and $\gamma\leq0$, where $\gamma=0$ yields the EAMSD $\langle y^2(t)\rangle\simeq\frac{2D}{\Gamma(1+\alpha)}t^\alpha$ of particles moving in static medium.
It can be found that the EAMSD depends on the power-law exponent $\gamma$ in the scale factor $a(t)$. If the medium expands fast enough with $\gamma>\alpha/2$, then the particle's diffusion behavior is enhanced from $t^\alpha$ to $t^{2\gamma}$. Otherwise, for $\gamma<\alpha/2$, even when the medium contracts with $\gamma<0$, the particle's intrinsic motion plays the dominating role and presents the subdiffusion behavior $t^\alpha$. It also can be found that the diffusion behavior $t^\alpha$ for power-law contracting medium is faster than $t^{\alpha-1}$  in Eq. \eqref{EAMSD1-y} for exponentially contracting medium.

For the ensemble-averaged TAMSD over the physical coordinate $y(t)$, we need to substitute the EAMSD in Eq. \eqref{MSD-1By} into Eq. \eqref{TAMSD}. For convenience, denote $\langle y^2(t)\rangle\simeq C_\beta t^{\beta}$ with $\beta=\max(2\gamma,\alpha)$ and $C_\beta$ being the diffusion coefficient in Eq. \eqref{MSD-1By}. Then for $\Delta\ll T$, it holds that
\begin{equation}\label{TAMSD-1By2}
\begin{split}
    \langle\overline{\delta^2(\Delta)}\rangle&\simeq \frac{C_\beta}{T-\Delta}\int_0^{T-\Delta} (t+\Delta)^\beta+t^\beta-2(t+\Delta)^\gamma t^{\beta-\gamma} dt\\
    &\simeq\left\{
    \begin{array}{ll}
          \frac{C_\beta \gamma^2}{2\gamma-1}T^{2\gamma-2}\Delta^2, &~ \gamma>\frac{\alpha}{2}, \\[5pt]
      \frac{C_\beta(\alpha-2\gamma)}{\alpha}T^{\alpha-1}\Delta, & ~\gamma<\frac{\alpha}{2}.
    \end{array}\right.
  \end{split}
\end{equation}
Similar to the EAMSD, the ensemble-averaged TAMSD also shows different diffusion behavior for particles moving in expanding medium with different rate $\gamma$. When the medium expands slowly with $0<\gamma<\alpha/2$ and contracts with $\gamma<0$, the particle's intrinsic diffusion plays the leading role, and presents the linear increasing with respect to the lag time $\Delta$. The linear increasing of ensemble-averaged TAMSD on $\Delta$ is common in large amount of diffusion processes, including (scaled) Brownian motion \cite{JeonChechkinMetzler:2014,ThielSokolov:2014}, subdiffusive CTRW \cite{LubelskiSokolovKlafter:2008,HeBurovMetzlerBarkai:2008}, heterogeneous diffusion processes \cite{CherstvyChechkinMetzler:2013,CherstvyMetzler:2013}, and random diffusivity processes \cite{WangChen:2021,WangChen:2022,HidalgoBarkaiBurov:2021}. When the medium expands fast with $\gamma>\alpha/2$ and plays the dominating role, however, the particle presents the ballistic behavior $\Delta^2$ for large $\Delta$, which is a quite interesting phenomena and only found in ballistic L\'{e}vy walk \cite{FroembergBarkai:2013,FroembergBarkai:2013-2} as well as its variants in external force fields \cite{ChenWangDeng:2019-3,ChenDeng:2021}.

\begin{figure}
\begin{minipage}{0.35\linewidth}
  \centerline{\includegraphics[scale=0.29]{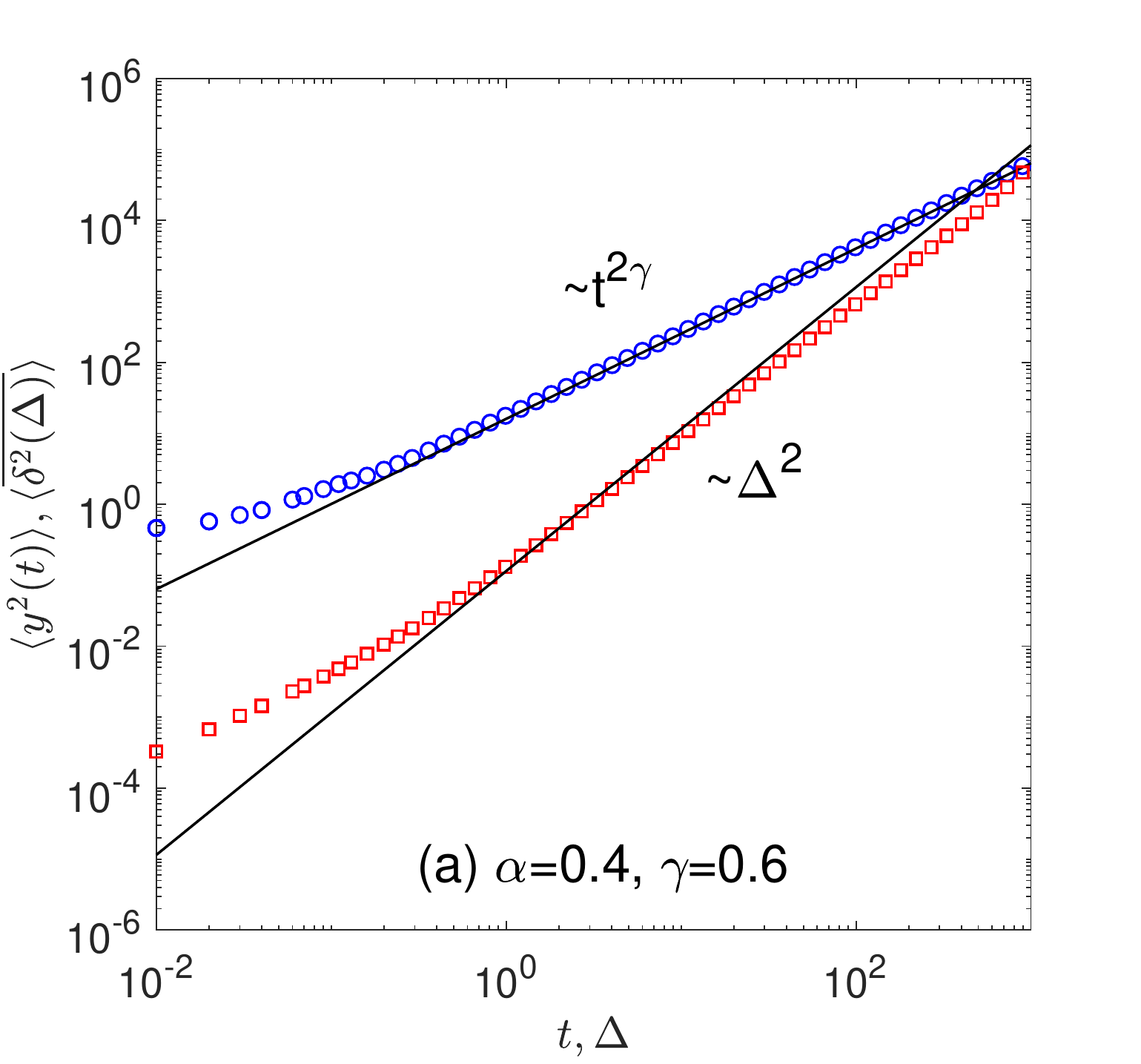}}
  \centerline{}
\end{minipage}
\hspace{1.43cm}
\begin{minipage}{0.35\linewidth}
  \centerline{\includegraphics[scale=0.29]{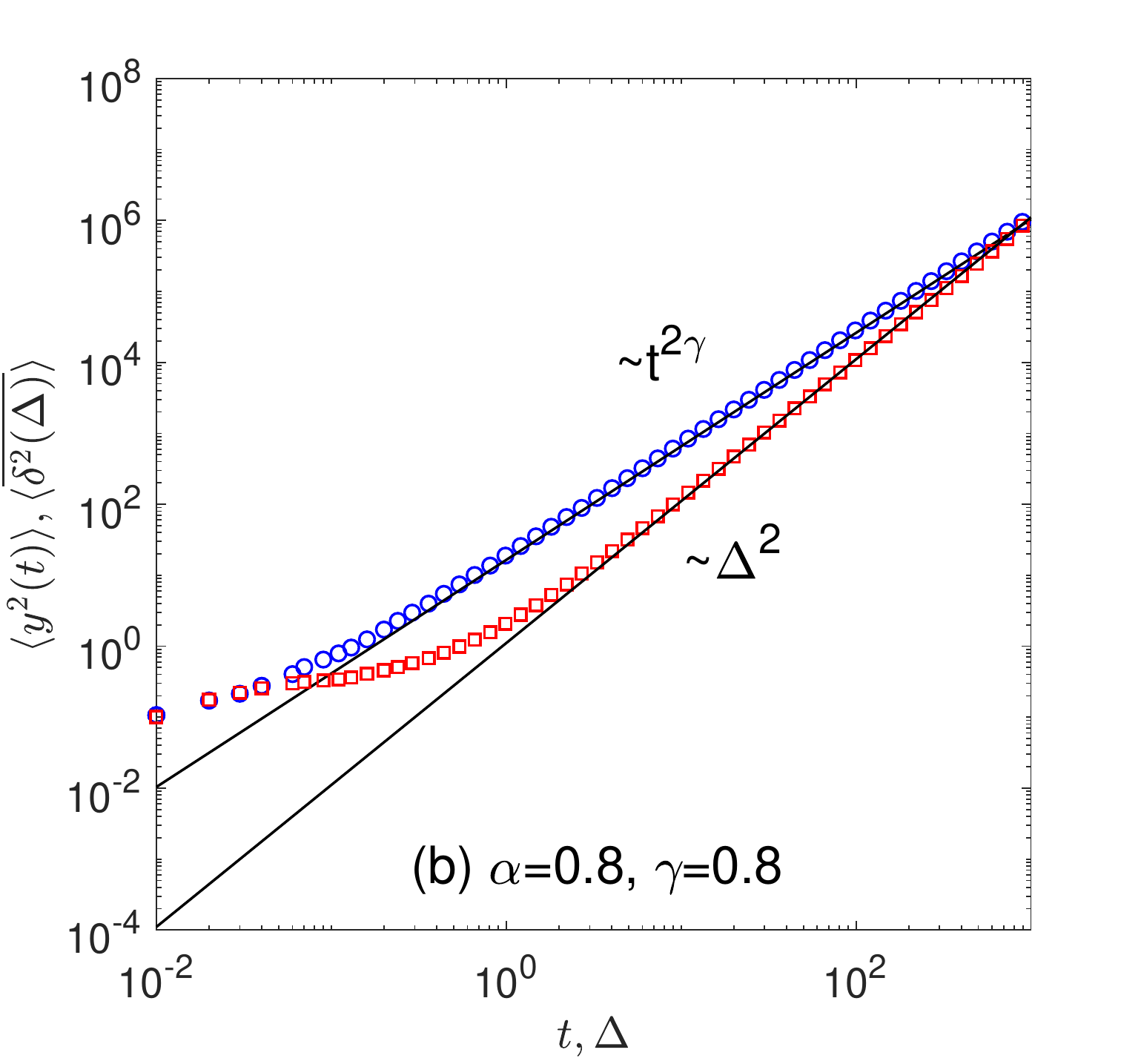}}
  \centerline{}
\end{minipage}
\begin{minipage}{0.35\linewidth}
  \centerline{\includegraphics[scale=0.29]{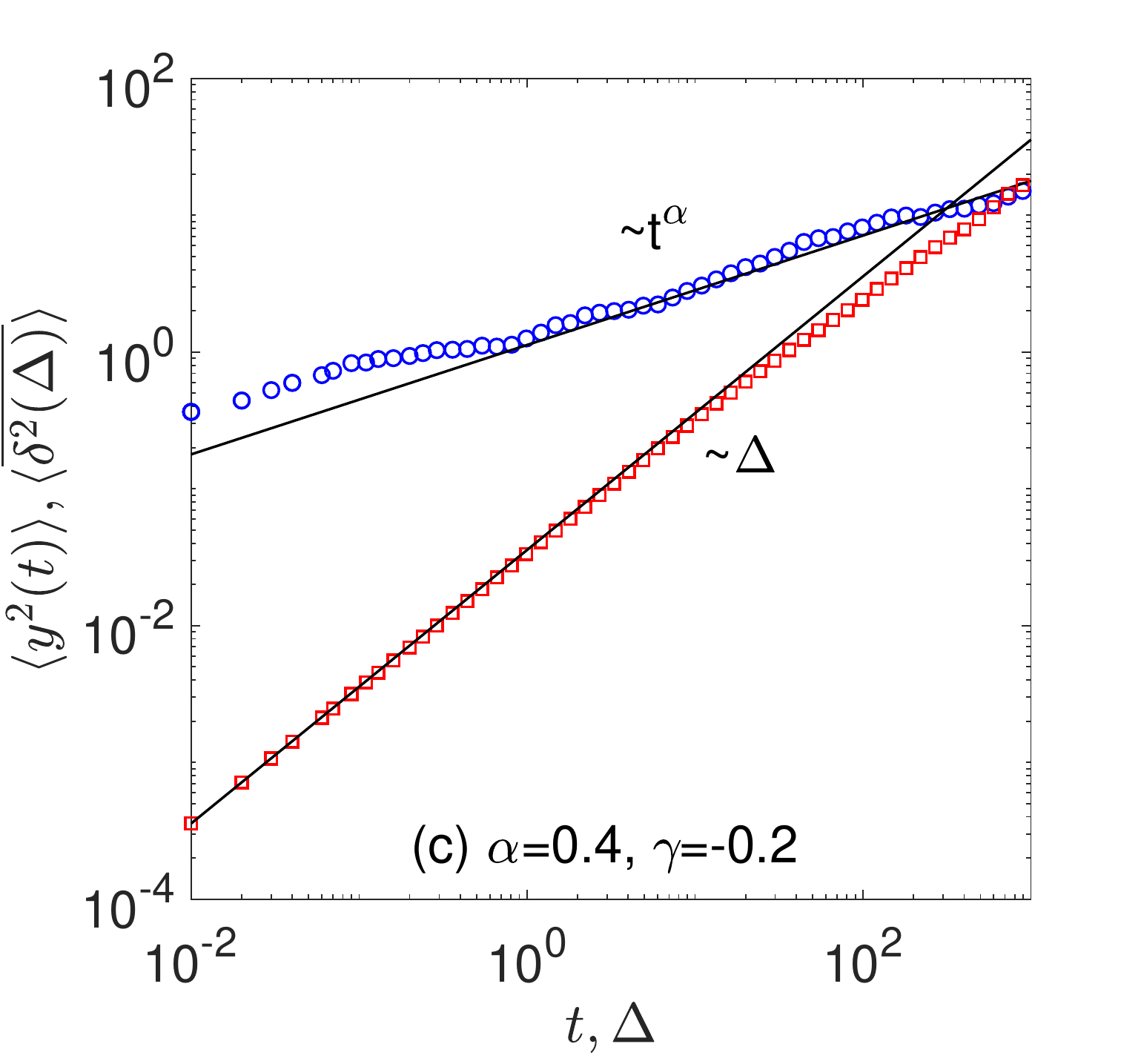}}
  \centerline{}
\end{minipage}
\hspace{1.43cm}
\begin{minipage}{0.35\linewidth}
  \centerline{\includegraphics[scale=0.29]{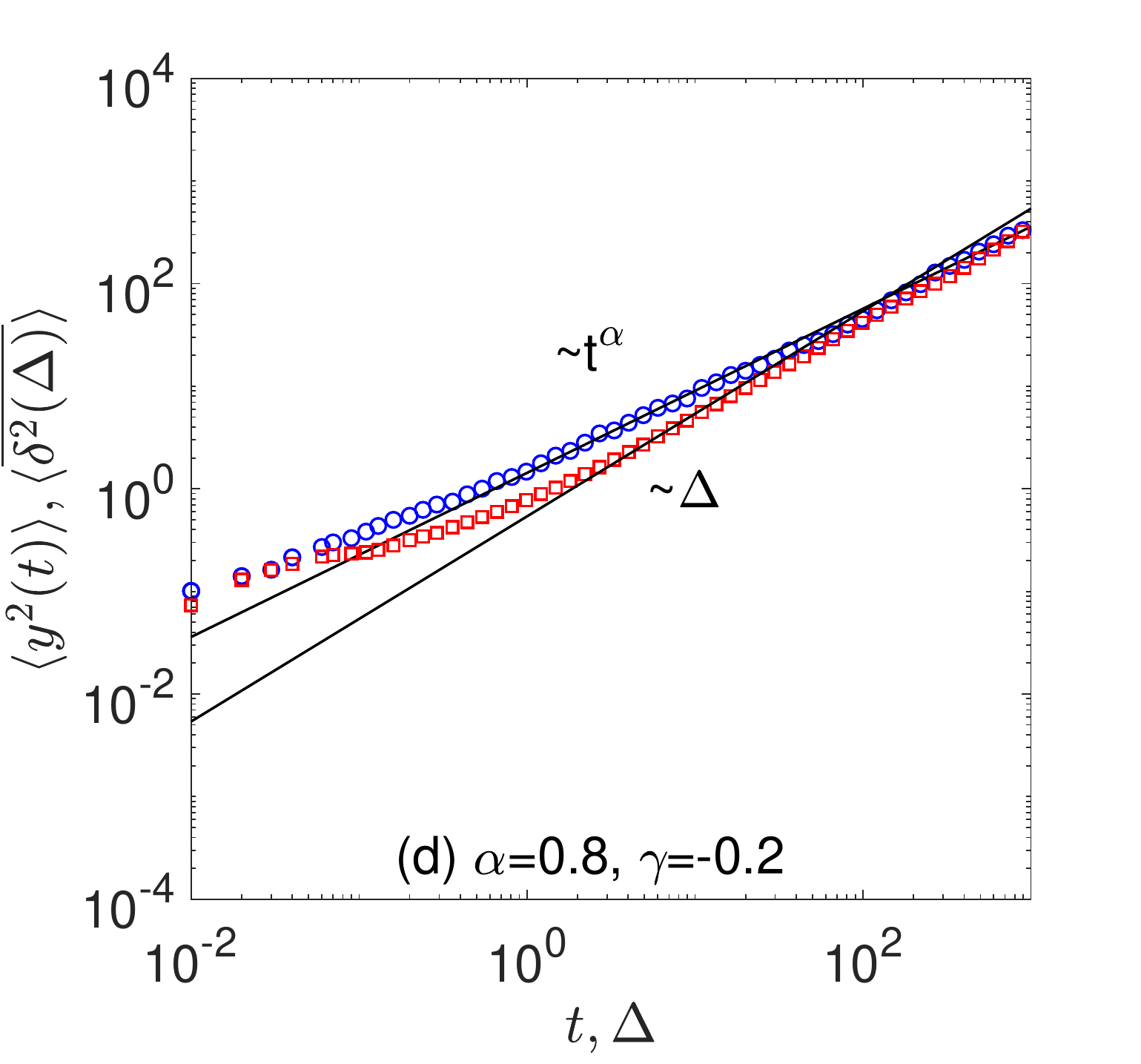}}
  \centerline{}
\end{minipage}
\caption{(Color online) EAMSD $\langle x^2(t)\rangle$ and ensemble-averaged TAMSD $\langle\overline{\delta^2(\Delta)}\rangle$ when the medium changes at the power-law rate with scale factor $a(t)=e^{\gamma t}$. The theoretical results for $\langle x^2(t)\rangle$ in Eq. \eqref{MSD-1By} and $\langle\overline{\delta^2(\Delta)}\rangle$ in Eq. \eqref{TAMSD-1By2} are shown by black solid lines. The blue circles and red squares denote the simulated EAMSD and ensemble-averaged TAMSD, respectively. They fit to the theoretical lines well in four panels with different $\alpha$ and $\gamma$. The last several simulation points for ensemble-averaged TAMSD deviate slightly from theoretical results due to the failure of condition $\Delta\ll T$.
Other parameters: $D=1$, $T=1000$, $t_0=0.1$, and the number of trajectories used for ensemble is $N=10^3$.}\label{fig2}
\end{figure}

The corresponding simulations are shown in Fig. \ref{fig2}, where we choose different $\alpha$ and $\gamma$. The positive $\gamma$, satisfying $\gamma>\alpha/2$ in the upper panel of Fig. \ref{fig2}, implies an expanding medium, while the negative one satisfying $\gamma<\alpha/2$ in the bottom panel implies a contracting medium. The difference between the EAMSD in Eq. \eqref{MSD-1By} and ensemble-averaged TAMSD in Eq. \eqref{TAMSD-1By2} (also in simulations) implies the non-ergodicity of particles moving in expanding medium with power-law-formed scale factor $a(t)=(\frac{t+t_0}{t_0})^\gamma$.

\section{Superdiffusion in expanding media}\label{Sec4}
Compared with the subordinated overdamped Langevin equation, the subordinated underdamped Langevin equation is often used to describe superdiffusion or finite-velocity processes \cite{HanSilvaKorabelFedotov:2021,GionaCairoliKlages:2022}.
Assume that the particle's intrinsic motion is characterized by the following set of Langevin equations \cite{EuleZaburdaevFriedrichGeisel:2012,WangChenDeng:2019}
\begin{equation}\label{model2}
    \dot{y}_I(t)=v(t), ~~~
    \dot{v}(s)=-\mu v(s) +\xi(s), ~~~
    \dot{t}(s)= \eta(s),
\end{equation}
where $y_I(t)$ denotes the particle's trajectory driven by its intrinsic motion, $v(s)$ is the particle's velocity in operational time $s$, $\mu$ is the friction coefficient, and $\xi(s)$ is the Gaussian white noise satisfying $\langle\xi(s_1)\xi(s_2)\rangle=2D\delta(s_1-s_2)$. Different from the subordinated overdamped Langevin equation in Sec. \ref{Sec3}, it is the velocity process in physical time $t$ here that is defined by $v(t):=v(s(t))$. The diffusion behavior and the ergodicity breaking of Eq. \eqref{model2} have been studied in Ref. \cite{WangChenDeng:2019}, which shows that L\'{e}vy-walk-like dynamic exhibits superdiffusion behavior \cite{EuleZaburdaevFriedrichGeisel:2012,WangChenDeng:2019,ChenWangDeng:2019-3}
\begin{equation}\label{MSD-2yi}
  \begin{split}
    \langle y_I^2(t)\rangle\simeq\left\{
    \begin{array}{ll}
          \frac{D(1-\alpha)}{\mu}t^{2}, &~ 0<\alpha<1, \\[5pt]
      \frac{2D\tau_0^{\alpha-1}(\alpha-1)}{\mu(2-\alpha)(3-\alpha)}t^{3-\alpha}, & ~1<\alpha<2.
    \end{array}\right.
  \end{split}
\end{equation}

Although the underdamped Langevin form in Eq. \eqref{model2} is more complex than the subdiffusion case in Sec. \ref{Sec3}, the relations in Eqs. \eqref{Eq-at2} and \eqref{Eq-at3} between physical coordinate and comoving coordinate are still valid. Therefore, the Langevin equation of the comoving coordinate $x(t)$ is
\begin{equation}\label{model2-x}
\begin{split}
 \dot{x}(t)=\frac{v(t)}{a(t)}, ~~~
    \dot{v}(s)=-\mu v(s) +\xi(s), ~~~
    \dot{t}(s)= \eta(s),
\end{split}
\end{equation}
for particles moving in the expanding medium with scale factor $a(t)$.
Based on the second equation in Eq. \eqref{model2-x}, the velocity correlation function in operational time $s$ can be obtained as
\begin{equation}
\langle v(s_1)v(s_2)\rangle= \frac{D}{\mu} (e^{-\mu |s_1- s_2|}-e^{-\mu(s_1+s_2)}).
\end{equation}
Then using the technique of subordination \cite{BauleFriedrich:2005,WangChenDeng:2019,ChenWangDeng:2019-3}, the velocity correlation function in physical time $t$ has the asymptotic behavior at long time (small $\lambda_1$ and $\lambda_2$):
\begin{equation}\label{model2-vvlambda}
\begin{split}
&\langle v(\lambda_1)v(\lambda_2)\rangle\\
&=\int_0^\infty \int_0^\infty\langle v(s_1)v(s_2)\rangle h(s_2,\lambda_2;s_1,\lambda_1)ds_1 ds_2\\
&\simeq\frac{D}{\mu}\frac{\Phi(\lambda_1)+\Phi(\lambda_2)-\Phi(\lambda_1+\lambda_2)}{\lambda_1\lambda_2\Phi(\lambda_1+\lambda_2)},
\end{split}
\end{equation}
where the expression of $h(s_2,\lambda_2;s_1,\lambda_1)$ is shown in Eq. \eqref{h2}.
Performing the inverse Laplace transform on Eq. \eqref{model2-vvlambda}, one obtains, for large $t_1$ and $t_2$ ($t_1<t_2$),
\begin{equation}\label{v1v2}
\begin{split}
&\langle v(t_1)v(t_2)\rangle\\
&\simeq \left\{
\begin{array}{ll}
  \frac{D\sin(\pi\alpha)}{\mu\pi}B\left(\frac{t_1}{t_2};\alpha,1-\alpha \right),  & ~0<\alpha<1,  \\[5pt]
  \frac{D\tau_0^{\alpha-1}}{\mu}\left((t_2-t_1)^{1-\alpha}-t_2^{1-\alpha}\right), &~ 1<\alpha<2,
\end{array}\right.
\end{split}
\end{equation}
where $B(x; a, b)$ is the incomplete Beta function.
Therefore, based on the first equation of Eq. \eqref{model2-x}, the position correlation function over comoving coordinate $x(t)$ can be obtained
\begin{equation}\label{corr-2x}
\begin{split}
\langle x(t_1)x(t_2)\rangle=\int_0^{t_1} \int_0^{t_2} \frac{\langle v(t'_1)v(t'_2)\rangle}{a(t'_1)a(t'_2)} dt'_1 dt'_2.
\end{split}
\end{equation}
Considering the relation $y(t)=a(t)x(t)$ between two coordinates at any time $t$, we also have the position correlation function in physical coordinate
\begin{equation}\label{corr-2y}
\begin{split}
\langle y(t_1)y(t_2)\rangle&=a(t_1)a(t_2)\int_0^{t_1} \int_0^{t_2} \frac{\langle v(t'_1)v(t'_2)\rangle}{a(t'_1)a(t'_2)} dt'_1 dt'_2,
\end{split}
\end{equation}
where the velocity correlation function is shown in Eq. \eqref{v1v2}.

\subsection{Scale factor $a(t)=e^{\gamma t}$}
\begin{figure}
  \centering
  \includegraphics[scale=0.55]{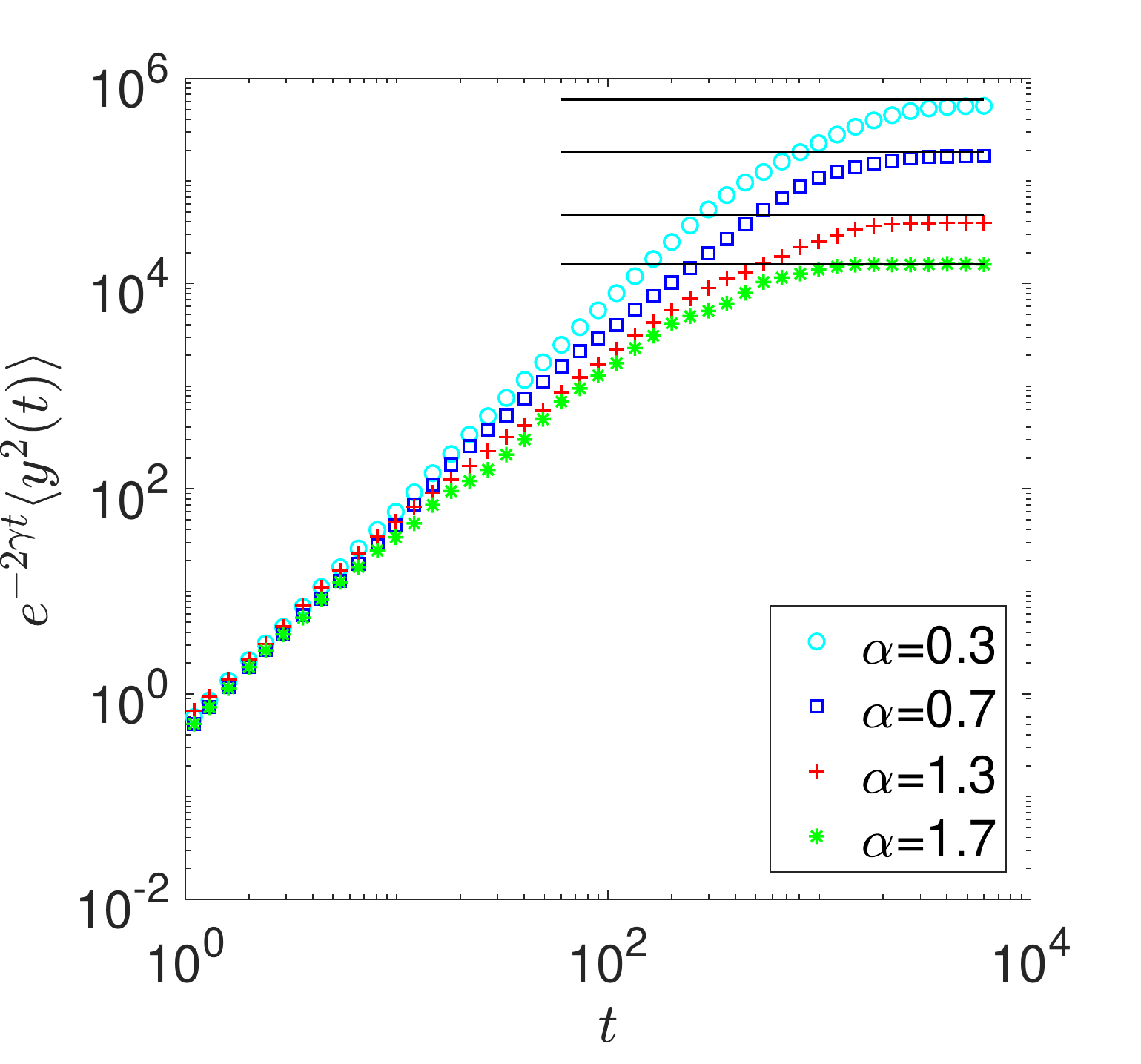}\\
  \caption{(Color online) EAMSDs $\langle x^2(t)\rangle$ of the comoving coordinate when the medium expands exponentially with scale factor $a(t)=e^{\gamma t}$. The $\langle x^2(t)\rangle$ tends to a constant $A\simeq e^{-2\gamma t}\langle y^2(t)\rangle$ in Eq. \eqref{const-A} and implies an exponentially increasing of $\langle y^2(t)\rangle$.
  The theoretical results for different $\alpha$ are shown by black solid lines. The markers (cyan circles, blue squares, red plus signs and green asterisks) denote the simulated EAMSDs with different $\alpha$, respectively.
Other parameters: $D=\mu=\tau_0=1$, $T=5000$, $\gamma=0.001$, and the number of trajectories used for ensemble is $N=10^3$.}\label{fig3}
\end{figure}

Similar to the discussions on subdiffusion in expanding medium in the previous section, the two cases of $\gamma>0$ and $\gamma<0$ will lead to different diffusion behaviors, and be discussed separately here.
For the exponentially expanding medium with $\gamma>0$, we have
\begin{equation}
\begin{split}
\langle x(t_1)x(t_2)\rangle=\int_0^{t_1}\int_0^{t_2}e^{-\gamma t_1'}e^{-\gamma t_2'}\langle v(t_1')v(t_2') \rangle dt_1'dt_2',
\end{split}
\end{equation}
the Laplace transform of which is
\begin{equation}\label{XCF-21}
\begin{split}
\langle x(\lambda_1)x(\lambda_2)\rangle&=\frac{1}{\lambda_1\lambda_2}\langle v(\lambda_1+\gamma)v(\lambda_2+\gamma)\rangle\\
&\simeq \frac{1}{\lambda_1\lambda_2}\langle v^2(\gamma)\rangle
\end{split}
\end{equation}
for small $\lambda_1$ and $\lambda_2$.
The expression of velocity correlation function in frequency domain is shown in Eq. \eqref{model2-vvlambda}. Then we perform the inverse Laplace transform on Eq. \eqref{XCF-21} and obtain
\begin{equation}
\begin{split}
\langle x(t_1)x(t_2)\rangle\simeq A:=\langle v^2(\gamma)\rangle,
\end{split}
\end{equation}
which tends to a constant $A$ at long time limit. Therefore, the EAMSD in physical coordinate is increasing exponentially:
\begin{equation}\label{EAMSD2-y}
\begin{split}
&\langle y^2(t)\rangle=a^2(t)\langle x^2(t)\rangle\simeq Ae^{2\gamma t},
\end{split}
\end{equation}
where
\begin{equation}\label{const-A}
\begin{split}
A=\left\{
\begin{array}{ll}
  \frac{D(2^{1-\alpha}-1)}{\mu\gamma^2},  & ~0<\alpha<1,  \\[5pt]
  \frac{D}{\mu}\frac{\Gamma(2-\alpha)(2^{\alpha-1}-1)\gamma^{\alpha-2}}{ \gamma-2^{\alpha-1}\Gamma(2-\alpha)\gamma^\alpha}, &~ 1<\alpha<2.
\end{array}\right.
\end{split}
\end{equation}
Note that the coefficient $A$ is obtained by use of the expression of $\Phi(\lambda)$ in Eq. \eqref{model2-vvlambda}, and the latter is approximatively given in Eq. \eqref{Phi} for small $\lambda$.
Thus, Eq. \eqref{const-A} is only valid for small $\gamma$. Although the accurate value of $A$ cannot be obtained for large $\gamma$, the sure thing is that $A$ is a constant and the EAMSD increases exponentially as Eq. \eqref{EAMSD2-y} shows. To get a good presentation in simulations, we show the EAMSDs $\langle x^2(t)\rangle$ of the comoving coordinate in Fig. \ref{fig3} for different $\alpha$. All the EAMSDs $\langle x^2(t)\rangle$ tend to a constant at long time limit, which is smaller for larger $\alpha$.

\begin{figure}
  \centering
  \includegraphics[scale=0.55]{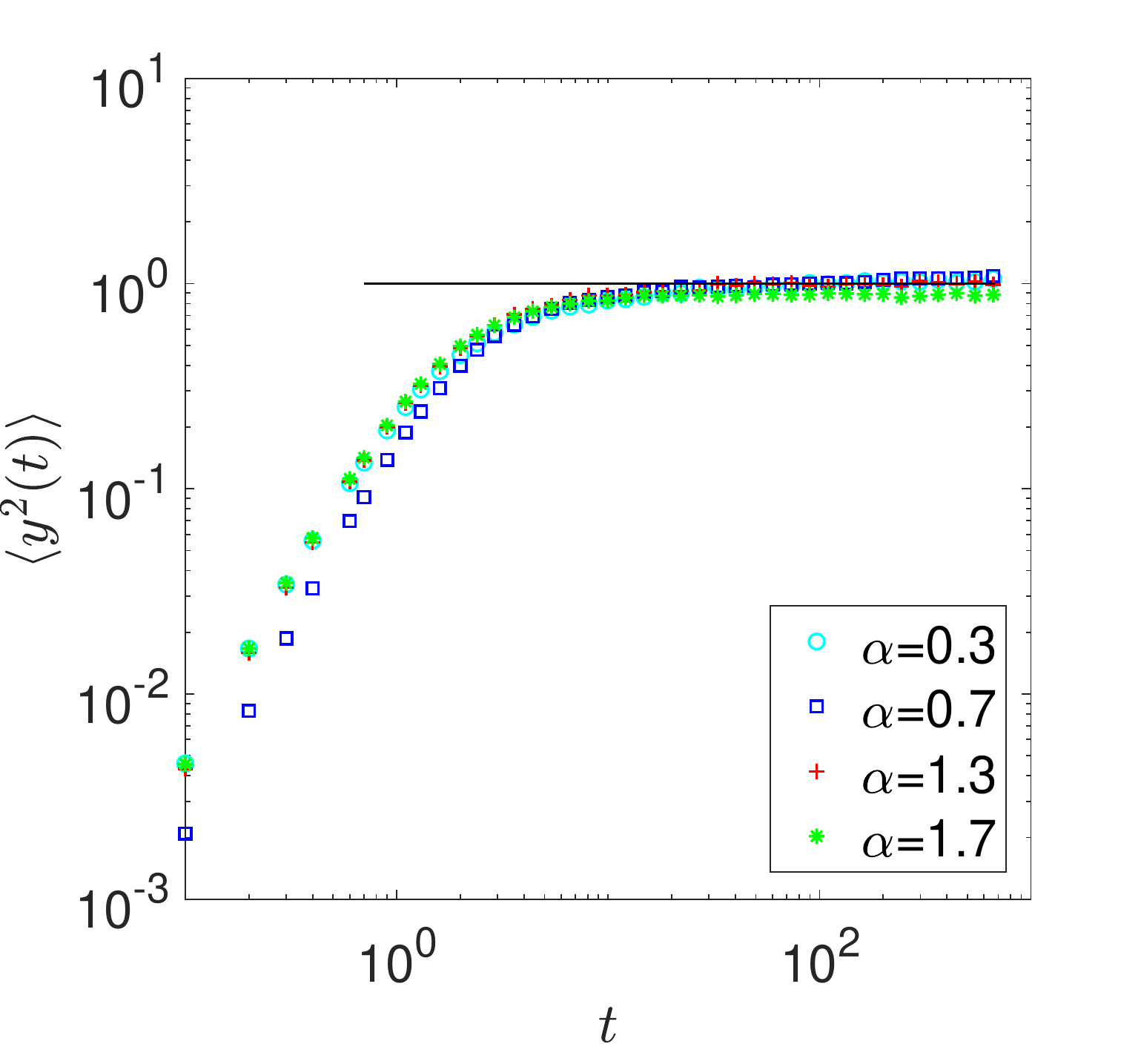}\\
  \caption{(Color online) EAMSDs $\langle y^2(t)\rangle$ of the physical coordinate when the medium contracts exponentially with scale factor $a(t)=e^{\gamma t}$.
  The theoretical results for different $\alpha$ are shown by the same black solid line. The markers (cyan circles, blue squares, red plus signs and green asterisks) denote the simulated EAMSDs with different $\alpha$, respectively.
Other parameters: $D=\mu=\tau_0=1$, $T=800$, $\gamma=-1$, and the number of trajectories used for ensemble is $N=10^3$.}\label{fig4}
\end{figure}

On the other hand, for contracting medium with $\gamma<0$, the corresponding position correlation function in physical coordinate is
\begin{equation}
\begin{split}
\langle y(t_1)y(t_2)\rangle=\int_0^{t_1}\!\!\!\int_0^{t_2} e^{\gamma(t_1-t_1')}e^{\gamma(t_2-t_2')}\langle v(t'_1)v(t'_2)\rangle dt'_1 dt'_2,
\end{split}
\end{equation}
which happens to be a convolution form. By using the technique of Laplace transform, we obtain the position correlation function in frequency domain:
\begin{equation}
\begin{split}
\langle y(\lambda_1)y(\lambda_2)\rangle&=\frac{1}{\lambda_1-\gamma}\frac{1}{\lambda_2-\gamma}\langle v(\lambda_1)v(\lambda_2)\rangle\\
&\simeq \frac{1}{\gamma^2}\langle v(\lambda_1)v(\lambda_2)\rangle,
\end{split}
\end{equation}
where we consider the asymptotic behavior for large $t_1$ and $t_2$ [i.e., for small $\lambda_1$ and $\lambda_2$]. Performing the inverse Laplace transform yields the EAMSD
\begin{equation}\label{EAMSD2-y2}
  \langle y^2(t)\rangle\simeq\frac{1}{\gamma^2}\langle v^2(t)\rangle\simeq \frac{D}{\mu \gamma^2},
\end{equation}
since the velocity process $v(t)$ in Langevin equation \eqref{model2-x} tends to a stationary state with variance $D/\mu$. In contrast to the $\alpha$-dependent result in Eq. \eqref{EAMSD2-y} for $\gamma>0$, the EAMSD $\langle y^2(t)\rangle$ of the physical coordinate tends to an $\alpha$-independent constant in Eq. \eqref{EAMSD2-y2}.
The corresponding simulations for different $\alpha$ are presented in Fig. \ref{fig4}. All the EAMSDs $\langle y^2(t)\rangle$ tend to the same constant at long time limit, which shows significant difference from the case $\gamma>0$ in Fig. \ref{fig3}.
Similar to the case of subdiffusion process discussed in Sec. \ref{Sec3}, the superdiffusive particles moving in exponentially contracting medium also show the same results as the case of particles influenced by a harmonic potential \cite{WangChenDeng:2020-2}.

\subsection{Scale factor $a(t)=(\frac{t+t_0}{t_0})^\gamma$}
\begin{figure}
  \centering
  \includegraphics[scale=0.55]{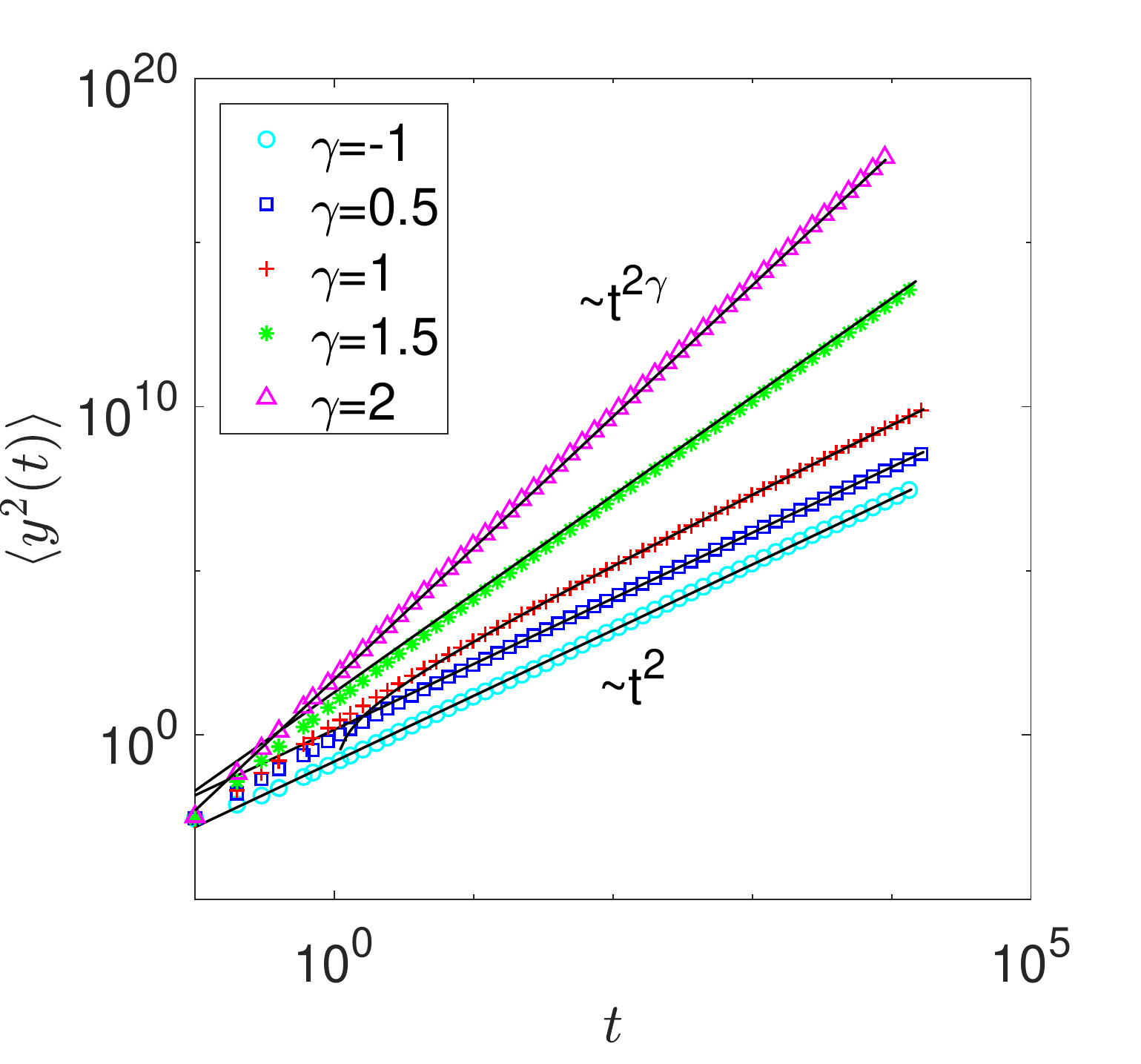}\\
  \caption{(Color online) EAMSD $\langle y^2(t)\rangle$ when $\alpha=0.5$ and the medium changes in a power-law rate with scale factor $a(t)=(1+t/t_0)^\gamma$.
  The theoretical results in Eq. \eqref{yy1} are shown by the black solid lines for different $\gamma$. The markers (cyan circles, blue squares, red plus signs, green asterisks and megenta triangles) denote the simulated EAMSDs with different $\gamma$, respectively.
Other parameters: $D=\mu=\tau_0=1$, $T=10^4$, $t_0=0.01$, and the number of trajectories used for ensemble is $N=10^3$.}\label{fig5}
\end{figure}

\begin{figure}
  \centering
  \includegraphics[scale=0.55]{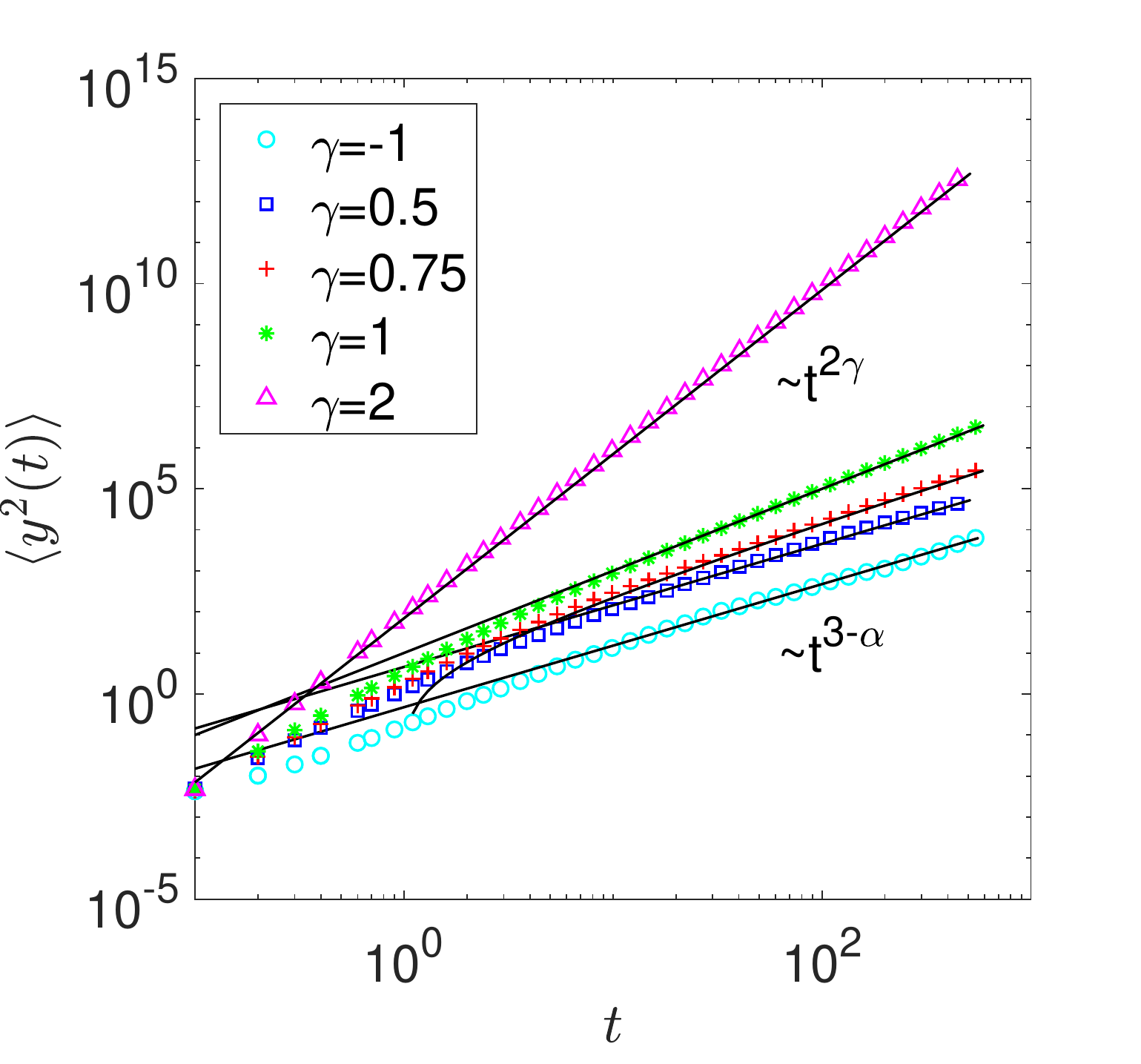}\\
  \caption{(Color online) EAMSD $\langle y^2(t)\rangle$ when $\alpha=1.5$ and the medium changes in a power-law rate with scale factor $a(t)=(1+t/t_0)^\gamma$.
  The theoretical results in Eq. \eqref{yy2} are shown by the black solid lines for different $\gamma$. The markers (cyan circles, blue squares, red plus signs, green asterisks and megenta triangles) denote the simulated EAMSDs with different $\gamma$, respectively.
Other parameters: $D=\mu=\tau_0=1$, $T=10^3$, $t_0=0.01$, and the number of trajectories used for ensemble is $N=10^3$.}\label{fig6}
\end{figure}

Now we focus on the case with the power-law-formed scale factor $a(t)=(\frac{t+t_0}{t_0})^\gamma$. Since the velocity correlation function shows discrepant asymptotic behaviors for $0<\alpha<1$ and $1<\alpha<2$ in Eq. \eqref{v1v2}, we analyze the particle's diffusion behavior in expanding medium separately. Due to the power-law form of $a(t)$, the calculation of the double integral in Eq. \eqref{corr-2y} is very complicated. Therefore, we present the final asymptotic behavior of EAMSD in physical coordinate here by putting the technical derivations and the explicit diffusion coefficients to Appendixes \ref{App4} and \ref{App5} for $0<\alpha<1$ and $1<\alpha<2$, respectively.
It holds that the EAMSD for $0<\alpha<1$ is
\begin{equation}\label{yy1}
  \begin{split}
    \langle y^2(t)\rangle\propto\left\{
    \begin{array}{ll}
          t^{2\gamma}, &~ \gamma>1, \\[5pt]
          t^2\ln t , &~ \gamma=1, \\[5pt]
          t^{2}, & ~\gamma<1,
    \end{array}\right.
  \end{split}
\end{equation}
and for $1<\alpha<2$ is
\begin{equation}\label{yy2}
\begin{split}
\langle y^2(t)\rangle
\propto \left\{
\begin{array}{ll}
 t^{2\gamma}, &~ \gamma>\frac{3-\alpha}{2},  \\[5pt]
 t^{3-\alpha}\ln t, &~\gamma=\frac{3-\alpha}{2},  \\[5pt]
 t^{3-\alpha},  & ~\gamma<\frac{3-\alpha}{2}.
\end{array}\right.
\end{split}
\end{equation}
The corresponding simulations are presented in Figs. \ref{fig5} and \ref{fig6} for $0<\alpha<1$ and $1<\alpha<2$, respectively.
For both cases, $\gamma=0$ yields the results of standard L\'{e}vy walk, i.e., the ballistic diffusion $t^2$ and sub-ballitic superdiffusion $t^{3-\alpha}$ for $0<\alpha<1$ and $1<\alpha<2$, respectively.
Similar to the result of subdiffusion case in Eq. \eqref{MSD-1By}, the diffusion behavior depends on the  relationship between the parameters $\gamma$ and $\alpha$. In the case of $0<\alpha<1$, the particle's intrinsic motion presents ballistic behavior $t^2$. If the medium expands fast enough with $\gamma>1$, then the diffusion behavior of particles is enhanced from $t^2$ to $t^{2\gamma}$. Otherwise, for $\gamma<1$, including the contracting medium with $\gamma<0$, the particle's intrinsic motion plays the leading role and presents the ballistic behavior $t^2$. The critical case with $\gamma=1$ shows an addition logarithmic increasing. While for the case of $1<\alpha<2$ with the particle's intrinsic motion presenting sub-ballistic superdiffusion behavior $t^{3-\alpha}$, the critical condition becomes $\gamma=\frac{3-\alpha}{2}$. Faster expanding rate with bigger $\gamma$ also leads to the superdiffusion $t^{2\gamma}$, and slower expanding rate with smaller $\gamma$ yields the intrinsic diffusion $t^{3-\alpha}$.
The biggest difference between the exponential and power-law-formed $a(t)$ is that when the medium contracts with $\gamma<0$, exponentially contracting medium changes the particle's intrinsic diffusion behavior while power-law contracting medium does not.

\section{Summary}\label{Sec5}

The dynamic mechanism of particles moving in expanding medium has been revealed in the framework of CTRW \cite{VotAbadYuste:2017,AngstmannHenryMcGann:2017,VotYuste:2018,Abad-etal:2020,VotAbadMetzlerYuste:2020}. To explore more anomalous diffusion processes in expanding medium and more physical observables, this paper proposes the Langevin picture of particle's trajectory in expanding medium. By using subordinated overdamped Langevin equation and underdamped Langevin equation to describe common subdiffusion and superdiffusion processes, respectively, we consider both exponential and power-law-formed scale factor $a(t)$, and find some interesting phenomena.

For the power-law-formed scale factor $a(t)$, there exists a critical value for the power law exponent of $a(t)$, a larger exponent implies a faster expanding rate of medium, and it enhances the particle's diffusion. While a smaller one implies a slower expanding rate or contracting medium, and it does not change particle's diffusion (See Eqs. \eqref{MSD-1By}, \eqref{TAMSD-1By2}, \eqref{yy1} and \eqref{yy2}). For the exponential formed $a(t)$, however, the medium changes fast, so that it produces a profound impact on particle's motion. The particle's diffusion behavior is enhanced to exponential form in exponentially expanding medium, but gets suppressed in exponentially contracting medium similar to the case with a harmonic potential (See Eqs. \eqref{EAMSD1-y}, \eqref{EATAMSD1-y}, \eqref{EAMSD2-y} and \eqref{EAMSD2-y2}).

In the subdiffusion case, the subordinated overdamped Langevin equation describes the same process as subdiffusive CTRW in scaling limit. Therefore,
by deriving the Fokker-Planck equation, evaluating the EAMSD, and comparing them to the results obtained in CTRW framework in Ref. \cite{VotAbadYuste:2017}, we verify the effectiveness of the Langevin approach proposed in this paper. Then we evaluate more physical observables, such as correlation function and TAMSD, and extend to the superdiffusion case. Since the subordinated underdamped Langevin equation models the L\'{e}vy-walk-like diffusion process in scaling limit, the discussions in the superdiffusion case reveal how L\'{e}vy-walk-like diffusion process behaves in expanding medium in some sense. More velocity-jump processes in expanding medium can be analyzed in the framework of (underdamped) Langevin equation as this paper shows.

The standard L\'{e}vy walk says that the particle moves on a straight line with a fixed speed for some random time \cite{ZaburdaevDenisovKlafter:2015}.
In fact, L\'{e}vy walk is not only analyzed in the form of a constant velocity, but also in a context of coupled CTRW, i.e., the so-called wait-first and jump-first models \cite{ZaburdaevDenisovKlafter:2015,MagdziarzZorawik:2017}. Instead of the constant velocity $v$ for duration time $\tau$ at one flight in standard L\'{e}vy walk, the wait-first L\'{e}vy walk says that the particle remains motionless for time $\tau$ and then executes a jump with length $v\tau$, resulting in a  discontinuous trajectory as the uncoupled CTRW. The jump-first L\'{e}vy walk differs from the wait-first case by the changed order of waiting and jumping moments.

Although the wait-first and jump-first models appear to be very similar to the standard L\'{e}vy walk, they have very different statistical properties, especially on the PDFs \cite{FroembergSchmiedebergBarkaiZaburdaev:2015}.
If we compare the three models only at the renewal moments where the direction of motion is chosen, we find that particle's positions are exactly the same. Thus the last renewal period plays a crucial role with regard to the differences between the three models. For EAMSD, it shows that the standard L\'{e}vy walk and wait-first one have the same diffusion behavior but with different diffusivity, while the jump-first one has diverging EAMSD \cite{MagdziarzZorawik:2017}.
Corresponding to the similarity between the wait-first L\'{e}vy walk and uncoupled CTRW, we take the same way to deal with the superdiffusion case as subdiffusion case. In other words, Sec. \ref{Sec3} provides the Langevin approach of analyzing the diffusion behavior of subdiffusive CTRW in expanding medium, while
Sec. \ref{Sec4} aims to investigate the diffusion behaviors of wait-first L\'{e}vy-walk-like processes in expanding medium.

\section*{Acknowledgments}
This work was supported by the National Natural Science Foundation of China under Grant No. 12105145 and No. 12205154, the Natural Science Foundation of Jiangsu Province under Grant No. BK20210325.

\appendix

\section{Derivation of Eq. \eqref{corr-newxi}}\label{App2}
We first introduce the Heaviside step function $\Theta(x)$, which satisfies $\Theta(x)=1$ for $x>0$, $\Theta(x)=0$ for $x<0$, and $\Theta(x=0)=1/2$. Then let $\langle B(s(t_1))B(s(t_2))\rangle$ be the correlation function of compound Brownian motion, where the brackets denote the ensemble averages on Brownian motion $B(\cdot)$ and inverse subordinator $s(t)$. Calculating the ensemble average on Brownian motion, we obtain
\begin{equation}\label{B1}
  \langle B(s(t_1))B(s(t_2))\rangle = \langle s(t_1)\Theta(t_2-t_1)+s(t_2)\Theta(t_1-t_2)\rangle.
\end{equation}
By using Eq. \eqref{Def-newnoise} and dividing $t_1$ and $t_2$ on both sides of Eq. \eqref{B1}, one arrives at
\begin{equation}
\begin{split}
\langle \bar\xi(t_1)\bar\xi(t_2)\rangle
&= \langle \dot{s}(t_1)\delta(t_2-t_1) \rangle \\
&= \delta(t_2-t_1)\frac{d}{dt_1}\langle s(t_1)\rangle\\
&= \frac{t_1^{\alpha-1}}{\Gamma(\alpha)}\delta(t_2-t_1),
\end{split}
\end{equation}
where the first moment of inverse subordinator \cite{BauleFriedrich:2005}
\begin{equation}
  \langle s(t_1)\rangle=\frac{t^\alpha}{\Gamma(\alpha+1)}
\end{equation}
has been used.

\section{Derivation of EAMSD in Eq. \eqref{yy1}}\label{App4}

When $0<\alpha<1$, substituting $a(t)=(\frac{t+t_0}{t_0})^\gamma$ and Eq. \eqref{v1v2} into Eq. \eqref{corr-2x}, we have
\begin{equation}\label{S4-1}
\begin{split}
\langle x^2(t)\rangle&\simeq \frac{2t_0^{2\gamma}D\sin(\pi\alpha)}{\mu\pi}\int_0^t \int_0^{t_2}(t_1+t_0)^{-\gamma}(t_2+t_0)^{-\gamma}  \\
&~~~~~~~~~~~~~~~\times B\left(\frac{t_1}{t_2},\alpha,1-\alpha\right)dt_1dt_2.
\end{split}
\end{equation}
It can be found that the value of $\gamma$ determines whether $\langle x^2(t)\rangle$ tends to a constant or infinity. More precisely, the internal integral grows at the rate of $t_2^{1-\gamma}$ with respect to $t_2$ as $t_2\rightarrow \infty$. Thus, the external integral [i.e., $\langle x^2(t)\rangle$] grows at the rate of $t^{2-2\gamma}$ with respect to $t$. Therefore, $\langle x^2(t)\rangle$ tends to a constant when $\gamma>1$ and to infinity when $\gamma\leq1$.
In detail, for $\gamma>1$, it holds that
\begin{equation}
\begin{split}
\langle y^2(t)\rangle=a^2(t)\langle x^2(t)\rangle\simeq C_1t^{2\gamma},
\end{split}
\end{equation}
where
\begin{equation*}
\begin{split}
C_1=\frac{2D\sin(\pi\alpha)}{\mu\pi}\int_0^\infty \int_0^{t_2} (t_1+t_0)^{-\gamma}(t_2+t_0)^{-\gamma} \\
\times
B\left(\frac{t_1}{t_2};\alpha,1-\alpha \right)dt_1dt_2.
\end{split}
\end{equation*}
While for $\gamma<1$, $\langle x^2(t)\rangle$ tends to infinity as $t\rightarrow\infty$, and thus, the long-time behavior of the integrand [i.e., large $t_1$ and $t_2$] plays the leading role. By considering this asymptotics, we solve the double integral in Eq. \eqref{S4-1}, and obtain
\begin{equation}
\begin{split}
\langle x^2(t)\rangle\simeq t_0^{2\gamma}C_2t^{2-2\gamma},
\end{split}
\end{equation}
and
\begin{equation}
\langle y^2(t)\rangle=a^2(t)\langle x^2(t)\rangle \simeq C_2t^{2},
\end{equation}
where
\begin{equation}
  C_2=\frac{D}{\mu(1-\gamma)^2}\left(1-\frac{\Gamma(1+\alpha-\gamma)}{\Gamma(\alpha)\Gamma(2-\gamma)}\right),
\end{equation}
which recovers the diffusion coefficient of standard L\'{e}vy walk in Eq. \eqref{MSD-2yi} by taking $\gamma=0$.
For the critical case $\gamma=1$, one arrives at $\langle y^2(t)\rangle\simeq C_3 t^2\ln t$. The diffusion coefficients $C_1$ and $C_3$ cannot be obtained explicitly due to the difficulty of evaluating the double integral in Eq. \eqref{S4-1}. For plotting the theoretical lines in simulations, we obtain the diffusion coefficients $C_1$ and $C_3$ by using the fitting method [see Fig. \ref{fig5}].

\section{Derivation of EAMSD in Eq. \eqref{yy2}}\label{App5}
When $1<\alpha<2$, substituting $a(t)=(\frac{t+t_0}{t_0})^\gamma$ and Eq. \eqref{v1v2} into Eq. \eqref{corr-2x}, we have
\begin{widetext}
  \begin{equation}\label{S5-1}
\begin{split}
\langle x^2(t)\rangle&\simeq2t_0^{2\gamma}\tau_0^{\alpha-1}\frac{D}{\mu}\int_0^t \int_0^{t_2}(t_1+t_0)^{-\gamma}
(t_2+t_0)^{-\gamma}((t_2-t_1)^{1-\alpha}-t_2^{1-\alpha})dt_1dt_2\\
&=2t_0^{3-\alpha}\tau_0^{\alpha-1}\frac{D}{\mu}\left[\frac{1}{2-\alpha}\int_0^{\frac{t}{t_0}} y^{2-\alpha}(y+1)^{-\gamma}{_2F_1}(1,\gamma,3-\alpha,-y)dy \right.\\
&~~~~~~\left.-\frac{1}{1-\gamma}\int_0^{\frac{t}{t_0}}((y+1)^{1-2\gamma}-(y+1)^{-\gamma})y^{1-\alpha}dy\right].
\end{split}
\end{equation}
\end{widetext}

Similar to Appendix \ref{App4}, we analyze the asymptotic behaviors of the two integrals above with respect to time $t$, and find two critical cases, which are $\gamma=(3-\alpha)/2$ and $\gamma=2-\alpha$. Therefore, we present the details for cases with different $\gamma$ in order.
For the case with $\gamma>\frac{3-\alpha}{2}$, the integral in Eq. \eqref{S5-1} converges as $t\rightarrow\infty$, and thus, the EAMSD in comoving coordinate $x$ tends to a constant, i.e., $\langle x^2(t)\rangle\simeq D_1$. However, we cannot obtain the exact value of $D_1$ due to the difficulty of evaluating the integral.

For the case $2-\alpha<\gamma<\frac{3-\alpha}{2}$, the integral of the three terms in Eq. \eqref{S5-1} can be solved, and we have
\begin{equation}\label{S5-2}
\begin{split}
\langle x^2(t)\rangle&\simeq 2t_0^{2\gamma}\tau_0^{\alpha-1}\frac{D}{\mu}(D_2 t^{3-2\gamma-\alpha} +D_3),
\end{split}
\end{equation}
where
\begin{equation}
  D_2=\frac{1}{3-2\gamma-\alpha}\left(\frac{\Gamma(1-\gamma)\Gamma(2-\alpha)}{\Gamma(3-\gamma-\alpha)}-
\frac{1}{1-\gamma}\right),
\end{equation}
and
\begin{equation}
\begin{split}
    D_3&=\frac{t_0^{3-2\gamma-\alpha}}{1-\gamma}\int_0^\infty (y+1)^{-\gamma}y^{1-\alpha}dy  \\
  &=\frac{t_0^{3-2\gamma-\alpha}}{1-\gamma}\frac{\Gamma(2-\alpha)\Gamma(\gamma+\alpha-2)}{\Gamma(\gamma)}.
\end{split}
\end{equation}
Since in this case, we have $3-2\gamma-\alpha>0$, the constant $D_3$ can be omitted.

Then for the case $\gamma<2-\alpha$, the EAMSD in comoving coordinate $x$ is
\begin{equation}\label{S5-5}
\begin{split}
\langle x^2(t)\rangle&\simeq 2t_0^{2\gamma}\tau_0^{\alpha-1}\frac{D}{\mu}
\Bigg( D_2t^{3-2\gamma-\alpha}  \\
&~~~~~~\left.+\frac{t_0^{1-\gamma}}{(2-\gamma-\alpha)(1-\gamma)}t^{2-\gamma-\alpha}\right),
\end{split}
\end{equation}
where the last term comes from the same integral as $D_3$ in Eq. \eqref{S5-2}. The difference is that $\gamma<2-\alpha$ makes the integral increase as $t\rightarrow\infty$. Considering $1<\alpha<2$, it holds that $3-2\gamma-\alpha>2-\gamma-\alpha$, and thus, the last term in Eq. \eqref{S5-5} can be also omitted as $D_3$ does.

For the critical case $\gamma=2-\alpha$, the EAMSD in comoving coordinate $x$ can be obtained as
\begin{equation}\label{S5-6}
\begin{split}
\langle x^2(t)\rangle&\simeq 2t_0^{4-2\alpha}\tau_0^{\alpha-1}\frac{D}{\mu}\left[\frac{\Gamma(\alpha)\Gamma(2-\alpha)-1}
{(\alpha-1)^2}t^{\alpha-1}\right. \\
&~~~~~~\left.+\frac{t_0^{\alpha-1}}{\alpha-1}\ln t\right],
\end{split}
\end{equation}
where $\ln t\ll t^{\alpha-1}$ and the last term can be omitted.
In fact, the leading terms in Eqs. \eqref{S5-2}, \eqref{S5-5} and \eqref{S5-6} are both
\begin{equation}
  \langle x^2(t)\rangle\simeq 2t_0^{2\gamma}\tau_0^{\alpha-1}\frac{D}{\mu}\cdot D_2t^{3-2\gamma-\alpha}.
\end{equation}
Therefore, the value range of $\gamma$ can be merged into $\gamma<\frac{3-\alpha}{2}$.

For another critical case $\gamma=\frac{3-\alpha}{2}$, the EAMSD in comoving coordinate $x$ is
\begin{equation}
\begin{split}
\langle x^2(t)\rangle\simeq  2t_0^{3-\alpha}\tau_0^{\alpha-1}\frac{D}{\mu}\cdot D_4\ln t,
\end{split}
\end{equation}
where
\begin{equation}
  D_4=\frac{\Gamma(\frac{\alpha-1}{2})\Gamma(2-\alpha)}{\Gamma(\frac{3-\alpha}{2})}-\frac{2}{\alpha-1}.
\end{equation}

In conclusion, the EAMSD in physical coordinate $y$ is
\begin{equation}
\begin{split}
&\langle y^2(t)\rangle=a^2(t)\langle x^2(t)\rangle\\
&\simeq \left\{
\begin{array}{ll}
  D_1t_0^{-2\gamma}t^{2\gamma}, &~ \frac{3-\alpha}{2}<\gamma,  \\[5pt]
  \frac{2\tau_0^{\alpha-1}DD_4}{\mu}t^{3-\alpha}\ln t, &~\gamma=\frac{3-\alpha}{2},  \\[5pt]
  \frac{2\tau_0^{\alpha-1}DD_2}{\mu}t^{3-\alpha},  & ~\gamma<\frac{3-\alpha}{2}.
\end{array}\right.
\end{split}
\end{equation}

\section*{References}
\bibliography{ReferenceW}

\end{document}